\documentclass[prl,aps,preprint,showpacs]{revtex4}%
\usepackage{amssymb}
\usepackage{amsmath}
\usepackage{amsfonts}
\usepackage{graphicx}%
\input{diagrams.tex}
\begin{document}
\author{A. Cabrera \ \ \&\ \ \ H. Montani}
\email{montani@cab.cnea.gov.ar}
\affiliation{Centro At\'{o}mico Bariloche and Instituto Balseiro, 8400 - S. C. de
Bariloche, R\'{\i}o Negro, Argentina}
\title{Hamiltonian Loop Group Actions and T-Duality for group manifolds}
\date{December 2004}

\begin{abstract}
We carry out a Hamiltonian analysis of Poisson-Lie T-duality based on the loop
geometry of the underlying phases spaces of the dual sigma and WZW models.
Duality is fully characterized by the existence of equivariant momentum maps
on the phase spaces such that the reduced phase space of the WZW model and a
pure central extension coadjoint orbit work as a bridge linking both the sigma
models. These momentum maps are associated to Hamiltonian actions of the loop
group of the Drinfeld double on both spaces and the duality transformations
are explicitly constructed in terms of these actions. Compatible dynamics
arise in a general collective form and the resulting Hamiltonian description
encodes all known aspects of this duality and its generalizations.

\end{abstract}
\maketitle

\section{Introduction}

Poisson-Lie T-duality \cite{KS-1} refers to a non-Abelian duality between two
$1+1$ $\sigma$-models describing the motion of a string on targets being a
dual pair of Poisson-Lie groups \cite{Drinfeld-1}, composing a perfect
Drinfeld double group \cite{Lu-We}. The Lagrangians of the models are written
in terms of the underlying bialgebra structure of the Lie groups, and
Poisson-Lie T-duality stems from the self dual character the Drinfeld double.
Classical T-duality transformation comes to relate some \emph{dualizable
}subspaces of these phase-spaces, mapping solutions reciprocally. It comes to
generalize the Abelian $R\longleftrightarrow R^{-1} $ \cite{Giveon} and
non-Abelian $G\longleftrightarrow\mathfrak{g}^{\ast}$ \cite{Lozano}%
,\cite{Alvarez-Liu} dualities which hold at classical and quantum level.
Former version appears tied to the dual symmetric structure of the target
manifolds of dual models \cite{Ivanov}. The Poisson Lie T-duality reproduces
all of them when the symplectic structure on the Drinfeld double $D=G\bowtie
G^{\ast}$ \cite{Alek-Malkin} is reduced to the cotangent bundle $T^{\ast}G$
with $G$ a trivial Poisson-Lie group.

A generating functional for PLT duality transformations \cite{KS-1}
\cite{Alvarez-npb} is constructed from the symplectic structure on $D$, and it
was shown \cite{Sfetsos} from algebraic properties in the dual Lagrangians
they are canonical ones (although their domains remain unclear). Also, for
closed string models, a Hamiltonian description \cite{Stern-1} reveals that
there exists Poisson maps from the \emph{T-dual} phase-spaces to the centrally
extended loop algebra of the Drinfeld double, and it holds for any hamiltonian
dynamics on this loop algebra and lifted to the T-dual phase-spaces.

In the pioneer works \cite{KS-1} \cite{Alek-Klim}, it was proposed a WZW-type
model with target on the Drinfeld double group $D$ from which a PL T-dual pair
of $\sigma$-models are obtained, providing a common roof and making clear how
PL T-duality works: solutions on a $\sigma$-model are lifted to the WZW model
on $D$ and then projected to the dual one. This setting makes natural the
appearance of the symplectic structure on $D$, in the generating functional of
the duality transformations. However, in contrast with the hamiltonian
approach in \cite{Stern-1}, the dynamic of the involved models were fixed to a
very particular form. It was also noted that PL T-duality just work on some
subspaces satisfying some \emph{dualizable} conditions expressed as monodromy
constraints. In the hamiltonian approach to the abelian $R\leftrightarrow
R^{-1}$ and non Abelian duality $G\leftrightarrow\mathfrak{g}^{\ast}$, the
dualizable spaces were well characterized \cite{Alvarez-Liu} leading, for
example, to the momentum-winding exchange, but it is unclear how to do the
same in PL case.

An approach in the framework of bicrossed product of Lie algebras is presented
in ref. \cite{Majid-Begg}, constructing and classifying many dual models for
the quasitriangular case, studying the possible orthogonal decomposition of
the Drinfeld double algebra and fixing appropriated hamiltonian dynamics.

The main aim of this work is to carry out a unified description of classical
PL T-duality based on the symplectic geometry of the loop groups spaces
involved in sigma and WZW models. We encode it in the commutative diagram%
\begin{equation}
\begin{diagram}[h=1.9em] &&& &(L{\mathfrak{d}}_{\Gamma }^{\ast};\;\{,\}_{KK})&&&&&\\ &&\ruTo^{\mu} && &&\luTo^{\tilde{\mu}} && &\\ (T^{\ast }LG;\omega _{o}) &&& &\uTo^{\hat{\Phi}}& &&& (T^{\ast }{LG^{\ast }};\;\tilde{\omega} _{o})&\\ &&\luTo &&&& \ruTo&& &\\ & &&& (\Omega D;\omega_{\Omega D} )&&& &&\\ \end{diagram} \label{T-duality-1}%
\end{equation}
where the left and right vertices are the phases spaces of the $\sigma
$-models, with the canonical Poisson (symplectic) structures, $L\mathfrak{d}%
_{\Gamma}^{\ast}$ is the dual of the centrally extended Lie algebra of $LD$
with the Kirillov-Kostant Poisson structure, and $\Omega D$ is the symplectic
manifold of based loops. In particular, alike $\hat{\Phi}$, we derive $\mu$
and $\tilde{\mu}$ as momentum maps associated to hamiltonian actions of the
centrally extended loop group $LD^{\wedge}$ on the $\sigma$-models. These
actions split the tangent bundles of the preimages under $\mu$ and $\tilde
{\mu}$ of the pure central extension orbit, and the dualizable subspaces are
identified as the orbits of $\Omega D$ which turn to be the symplectic
foliation. We shall show that the restriction of the diagram to these
subspaces, with symplectic arrows, gives precise description of the PL
T-duality embodying its essential features and providing a clear framework to
link with other approaches. From this setting, we shall be able to build dual
hamiltonian models by taking any suitable hamiltonian function on the loop
algebra of the double and lifting it in a collective form \cite{Guill-Sten}.
For particular choices, the lagrangian formalism will be reconstructed
obtaining the known dual $\sigma$-models and the master WZW-like model
encoding them.

This work is organized as follows: in Section I, we review the main features
of the symplectic geometry of the WZW model; in Section II, we describe the
actions of the $LD^{\wedge}$ on the phase spaces of the sigma model with
target $G$ and $G^{\ast}$, constructing the associated momentum maps and
explaining the connection of the group of based loops with this phase spaces
of the sigma models; in Section III, the contents of the diagram $\left(
\ref{T-duality-1}\right)  $ are developed, presenting the geometric
description of the PLT-duality. The dynamical questions are addressed in
Section IV, dealing with the hamiltonian and lagrangian descriptions of the
PLT-dual models. In Section V, we illustrate the construction for the Abelian
and non-Abelian duality, giving the explicit duality transformations and
identifying the dualizable subspaces. Finally, some conclusion and comments
are condensed in the last Section.

\section{I- The WZW model phase space geometry}

The WZW model was proposed by Witten \cite{Witten} as a modification of the
principal sigma model driving to equation of motion admitting factorizable
general solutions: $g\left(  \sigma,t\right)  =g_{l}\left(  \sigma+t\right)
g_{r}\left(  \sigma-t\right)  $ or $g\left(  \sigma,t\right)  =g_{r}\left(
\sigma-t\right)  g_{l}\left(  \sigma+t\right)  .$This is attained by adding to
the original action of the sigma model the Wess-Zumino term, and the order of
the light cone factors in $g\left(  \sigma,t\right)  $ depends on the sign of
the added term.

As it is well known, the phase space of a sigma model with target space the
group manifold $G$ is the cotangent bundle $T^{\ast}LG$ of the loop group $LG$
that turns to be a symplectic manifold with the canonical symplectic form
$\omega_{o}$ \cite{Abr-Mars}, and the dynamics is determined by the election
of the Hamiltonian function. However, there is no election of Hamiltonian
function on $\left(  T^{\ast}LG,\omega_{o}\right)  $ driving to equations of
motion equivalent to the WZW ones. In fact, as shown in ref. \cite{Fad-Takh},
the addition of Wess-Zumino term amounts to a modification of the canonical
Poisson brackets on $T^{\ast}LG$. It symplectic counterpart is exhaustively
studied in references \cite{Harnad-1},\cite{Feheretal} and references therein,
where a cocycle extension of the canonical symplectic form $\omega_{o}$ is
considered in combination with the Marsden-Weinstein reduction procedure in
order to recover the WZW equation of motion. This last description provides
the framework for our approach to Poisson-Lie T-duality, so it is worthwhile
to briefly review it below.

Let us consider a connected Lie group $D$ and its loop group $LD$. For $l\in
LD$, $l^{\prime}$ denotes the derivative in the loop parameter $\sigma\in
S^{1}$, and we write $vl^{-1}$ and $l^{-1}v$ for the right and left
translation of any vector field $v\in TD$. Let $\mathfrak{d}$ be the Lie
algebra of $D$ equipped with a non degenerate symmetric $Ad$-invariant
bilinear form $\left(  ,\right)  _{\mathfrak{d}}$. Frequently we will work
with the subset $L\mathfrak{d}^{\ast}\subset$ $(L\mathfrak{d})^{\ast}$ instead
of $(L\mathfrak{d})^{\ast}$ itself, and we identify it with $L\mathfrak{d}$
through the map $\psi:L\mathfrak{d}\rightarrow L\mathfrak{d}^{\ast}$ provided
by the bilinear form
\[
\left(  ,\right)  _{L\mathfrak{d}}\equiv\dfrac{1}{2\pi}\int_{S^{1}}\left(
,\right)  _{\mathfrak{d}}%
\]
on $L\mathfrak{d}$. This bilinear form defines a $2$-cocycle $\Gamma
_{k}:L\mathfrak{d}\times L\mathfrak{d}\rightarrow\mathbb{R}$ \cite{Pres-Seg},
\[
\Gamma_{k}(X,Y)=\frac{k}{2\pi}\int_{S^{1}}\left(  X\left(  \sigma\right)
,Y^{\prime}\left(  \sigma\right)  \right)  _{\mathfrak{d}}\,d\sigma
\]
with $X,Y\in L\mathfrak{d}$. It is invariant under the action of the
orientation preserving diffeomorphism group of the circle, $Diff^{+}(S^{1})$,
and invariant under the adjoint action of constant loops, $\Gamma
_{k}(Ad_{l_{o}}X,Ad_{l_{o}}Y)=\Gamma_{k}(X,Y)$, for $l_{o}\in D$. It can be
derived from the $1$-cocycle $C_{k}:LD\rightarrow L\mathfrak{d}^{\ast}$,
\[
C_{k}\left(  l\right)  =k\psi\left(  l^{\prime}l^{-1}\right)  ~~.
\]

We identify the cotangent bundle $T^{\ast}LD$ with $LD\times\left(
L\mathfrak{d}\right)  ^{\ast}$ by left translation and, in practice, we shall
work on $L\left(  D\times\mathfrak{d}^{\ast}\right)  $. The pair $\left(
T^{\ast}LD,\omega_{o}\right)  $, where $\omega_{o}$ is the canonical $2$-form
defined as \cite{Abr-Mars}%
\[
\mathbf{\langle}\omega_{o},(v,\rho)\otimes(w,\xi)\mathbf{\rangle}%
_{(l,\varphi)}=-\mathbf{\langle}\rho,l^{-1}w\mathbf{\rangle}_{L\mathfrak{d}%
}+\mathbf{\langle}\xi,l^{-1}v\mathbf{\rangle}_{L\mathfrak{d}}+\mathbf{\langle
}\varphi,[l^{-1}v,l^{-1}w]\mathbf{\rangle}_{L\mathfrak{d}}%
\]
for $(v,\rho),(w,\lambda)\in T_{(l,\varphi)}L\left(  D\times\mathfrak{d}%
^{\ast}\right)  $, is the symplectic manifold on which sigma models with
targets $D$ are framed on. As explained above, the WZW model doesn't fit this
symplectic structure. Indeed, the symplectic manifold underlying the chiral
sectors of the WZW model is $\left(  T^{\ast}LD,\omega_{\Gamma}\right)  $,
with $\omega_{\Gamma}$ being a symplectic $2$-form obtained by adding a
cocycle term to $\omega_{o}$ ,
\[
\mathbf{\langle}\omega_{\Gamma},\lambda_{\ast}^{-1}(v,\rho)\otimes
\lambda_{\ast}^{-1}(w,\xi)\mathbf{\rangle}_{(l,\varphi)}=\mathbf{\langle
}\omega_{o},(v,\rho)\otimes(w,\xi)\mathbf{\rangle}_{(l,\varphi)}-\Gamma
_{k}\left(  vl^{-1}\,,wl^{-1}\right)
\]
for $(v,\rho),(w,\lambda)\in T_{(l,\varphi)}L\left(  D\times\mathfrak{d}%
^{\ast}\right)  $. This symplectic structure has also a natural interpretation
in terms of symplectic groupoids \cite{Grupoids} for the underlying infinite
dimensional affine Poisson algebra (for details see \cite{Loops}). Indeed, the
cocycle $\Gamma_{k}$ defines an Affine Poisson structure on $L\mathfrak{d}%
^{\ast}$ induced by the action groupoid $H=LD\rhd L\mathfrak{d}^{\ast
}\rightrightarrows L\mathfrak{d}_{Aff}^{\ast}, $ with $LD$ acting by the
(right) affine coadjoint action $A_{l}(\xi)=Ad_{l}^{\ast}\xi+C_{k}\left(
l^{-1}\right)  ,$ and supplied with the symplectic form
\[
\omega_{\Gamma}^{R}=\omega_{o}^{R}-\Gamma_{k}\left(  dll^{-1}\overset{\otimes
}{,}dll^{-1}\right)
\]
where $\omega_{o}^{R}$ is the symplectic form on $L\left(  D\times
\mathfrak{d}^{\ast}\right)  $ obtained from the standard one on $T^{\ast}LD$
trivialized by right translations. So we see that $\omega_{\Gamma}^{R}$
becomes the above introduced $\omega_{\Gamma}$ under the diffeomorphism
$(l,\varphi)\longrightarrow(l,Ad_{l^{-1}}^{\ast}\varphi)$ which switches from
left to right trivialization of $T^{\ast}LD.$

Observe that $\omega_{\Gamma}$ it is no longer a bi-invariant $2$-form, only
the invariance under right translation $\varrho_{m}^{L\left(  D\times
\mathfrak{d}^{\ast}\right)  }(l,\mu)=(lm^{-1},Ad_{m^{-1}}^{\ast}\mu)$, $m\in
LD$, remains. Tied to it there is a non $Ad$-equivariant momentum map
$J^{R}:L\left(  D\times\mathfrak{d}^{\ast}\right)  \rightarrow L\mathfrak{d}%
^{\ast}$\textit{, }
\[
J^{R}\left(  l,\xi\right)  =-\xi+k\psi\left(  l^{-1}l^{\prime}\right)
\]
with associated $1$-cocycle $-C_{k}$, so that $J^{R}\left(  \varrho
_{m}^{L\left(  D\times\mathfrak{d}^{\ast}\right)  }\left(  l,\xi\right)
\right)  -Ad_{m^{-1}}^{\ast}J^{R}\left(  l,\xi\right)  =-k\psi\left(
k^{\prime}k^{-1}\right)  $ and $J^{R}$ is a Poisson map to $L\mathfrak{d}%
_{Aff}^{\ast}$ .

When the corresponding central extension $LD^{\wedge}$ of $LD$ does exist,
$\omega_{\Gamma}$ can be obtained from the standard symplectic structure on
$T^{\ast}LD^{\wedge}\overset{Left}{\simeq}LD^{\wedge}\times(L\mathfrak{d}%
_{\Gamma})^{\ast}$ by reduction under the corresponding $S^{1}\subset
LD^{\wedge}$ action and the $Ad$-equivariance of $J^{R}$ is then restored
substituting $L\mathfrak{d}$ by the centrally extended Lie algebra
$L\mathfrak{d}_{\Gamma}$, defined by the cocycle $\Gamma_{k}$. The centrally
extended adjoint and coadjoint actions of $LD^{\wedge}$ on $L\mathfrak{d}%
_{\Gamma}$ and $L\mathfrak{d}_{\Gamma}^{\ast}$ are defined as
\begin{align*}
\widehat{Ad}_{l}\left(  X,a\right)   &  =\left(  Ad_{l}X\,,\,a+k\left\langle
\psi\left(  l^{\prime}l^{-1}\right)  ,X\right\rangle \right) \\
\widehat{Ad}_{l^{-1}}^{\ast}\left(  \xi,b\right)   &  =\left(  Ad_{l^{-1}%
}^{\ast}\xi+bk\psi\left(  l^{\prime}l^{-1}\right)  \,,\,b\right)
\end{align*}
Note that the $S^{1}\subset LD^{\wedge}$ action is trivial and the embedding
$\xi\hookrightarrow($ $\xi,1)$ is a Poisson map from $L\mathfrak{d}%
_{Aff}^{\ast}$ to $L\mathfrak{d}_{\Gamma}\sim L\mathfrak{d}\times\mathbb{R}$
which maps the \emph{affine} coadjoint action of $LD$ to the centrally
extended one of $LD\hookrightarrow LD^{\wedge}$. Now, the extended momentum
map $\hat{J}^{R}:L\left(  D\times\mathfrak{d}^{\ast}\right)  \longrightarrow
L\mathfrak{d}_{\Gamma}^{\ast}$ is
\begin{equation}
\hat{J}^{R}\left(  l,\xi\right)  =\left(  J^{R}\left(  l,\xi\right)
,1\right)  =\left(  k\psi\left(  l^{-1}l^{\prime}\right)  -\xi\,,1\right)
\label{J-R}%
\end{equation}
which is $\widehat{Ad}^{LD}$-equivariant, $\hat{J}^{R}\left(  \varrho
_{m}^{L\left(  D\times\mathfrak{d}^{\ast}\right)  }\left(  l,\xi\right)
\right)  -\widehat{Ad}_{m^{-1}}^{LD\ast}\hat{J}^{R}\left(  l,\xi\right)
=\left(  0,0\right)  $.

The next step is to apply the Marsden-Weinstein reduction procedure
\cite{Mars-Wein} to the point $\left(  0,1\right)  $ $\in L\mathfrak{d}%
_{\Gamma}^{\ast}$. The restriction of $\omega_{\Gamma}$ to $\left(  \hat
{J}^{R}\right)  ^{-1}\left(  0,1\right)  $ defines the degenerate $2$-form
\begin{equation}
\gamma(v,w)=\Gamma(l^{-1}v,l^{-1}w) \label{gamma}%
\end{equation}
with null distribution generated by the infinitesimal action of constant loops
$D$. In fact, in order to obtain the reduced space, $\left(  \hat{J}%
^{R}\right)  ^{-1}\left(  0,1\right)  $ must be quotiented by the stabilizer
of $\left(  0,1\right)  $ $\in L\mathfrak{d}_{\Gamma}^{\ast}$, that is, by the
subgroup of constant loops, $Stab\left(  0,1\right)  =D$. Since $\left(
\hat{J}^{R}\right)  ^{-1}\left(  0,1\right)  =\left\{  \left(  l,k\psi\left(
l^{-1}l^{\prime}\right)  \right)  \,/\,l\in LD\right\}  \cong LD$, the reduced
space can be identified with the subgroup of \emph{based loops }
\[
\dfrac{\left(  \hat{J}^{R}\right)  ^{-1}(0)}{D}\equiv\Omega D=\left\{  \left[
l\right]  =ll^{-1}\left(  0\right)  ~/~l\in LD\right\}  ~.
\]
so that the fibration
\begin{equation}
\Lambda:LD\rightarrow\Omega D~~/~~~\Lambda\,\left(  l\right)  =\left[
l\right]  ~~~, \label{Lambda}%
\end{equation}
with fiber $D$, provides the symplectic $2$-form $\omega_{\Omega D}$ on
$\Omega D$ defined from $\Lambda^{\ast}\omega_{\Omega D}=\gamma$.

After the reduction procedure, $\omega_{\Omega D}$\ is still invariant under
the\ residual left action of $LD$\ on $\Omega D$
\begin{equation}
LD\times\Omega D\rightarrow\Omega D~~~/~~~\left(  m,\left[  l\right]  \right)
\longrightarrow\left[  ml\right]  \label{res-l-inv}%
\end{equation}
The associated momentum map $\Phi:\Omega D\rightarrow L\mathfrak{d}^{\ast
}\,/\,\,\Phi\left(  \left[  l\right]  \right)  =k\psi\left(  l^{\prime}%
l^{-1}\right)  $ is not $Ad$-equivariant. Again, introducing the extended
momentum map
\begin{equation}
\hat{\Phi}:\Omega D\rightarrow L\mathfrak{d}_{\Gamma}^{\ast}~~/~\hat{\Phi
}\left(  \left[  l\right]  \right)  =(\Phi\left(  \left[  l\right]  \right)
,\;1)=\widehat{Ad}_{l^{-1}}^{LD\ast}\left(  0,1\right)  ~~~, \label{phi-ext}%
\end{equation}
the equivariance is restored and the vertical arrow in diagram $\left(
\ref{T-duality-1}\right)  $ is explained. In fact, remember that
$L\mathfrak{d}_{\Gamma}^{\ast}$ is a Poisson manifold with Poisson bracket
$\left\{  f,g\right\}  _{KK}\left(  \eta\right)  =\left\langle \eta,\left[
df,dg\right]  _{L\mathfrak{d}_{\Gamma}}\right\rangle $ and their symplectic
leaves are the coadjoint orbits. Thereby, $\hat{\Phi}$ becomes into a
symplectic map (local diffeomorphism) onto the pure central extension orbit
$\mathcal{O}_{(0,1)}\equiv\mathcal{O}$, $\hat{\Phi}:(\Omega D,$ $\omega
_{\Omega D})\rightarrow$ $(\mathcal{O},\omega_{KK})$, with $\omega_{KK}%
$\textit{\ }being the Kirillov-Kostant symplectic form that on the
$\mathcal{O}$ reduces to%
\[
\left\langle \omega_{KK},\widehat{ad}_{X}^{\ast}(k\psi\left(  l^{\prime}%
l^{-1}\right)  ,1)\otimes\widehat{ad}_{Y}^{\ast}(k\psi\left(  l^{\prime}%
l^{-1}\right)  ,1))\right\rangle _{(k\psi\left(  l^{\prime}l^{-1}\right)
,1)}=\Gamma_{k}(X,Y)
\]
for $X,Y\in\left(  L\mathfrak{d}/\mathfrak{d}\right)  _{\Gamma}$. Then, for
any vector $\left[  v\right]  \in T_{\left[  l\right]  }\Omega D,$one has
$\left(  \hat{\Phi}\right)  _{\ast}\left[  v\right]  =-ad_{vl^{-1}}^{\ast
}Ad_{l^{-1}}^{\ast}\left(  0,1\right)  $ and
\[
\left\langle \left(  \hat{\Phi}\right)  ^{\ast}\omega_{KK},\left[  v\right]
\otimes\left[  w\right]  \right\rangle _{\left[  l\right]  }=\Gamma
(l^{-1}v,l^{-1}w)=\left\langle \omega_{\Omega D},\left[  v\right]
\otimes\left[  w\right]  \right\rangle
\]
It is worth remarking that only the coadjoint orbit through the pure central
extension element $(0,a)$, namely $\left(  \mathcal{O}_{(0,a)},\omega
_{KK}\right)  $, is (locally) symplectomorphic to $(\Omega D,$ $\omega_{\Omega
D})$.

\section{II- Hamiltonian $LD$ actions on dual phase-spaces}

In the following subsections we shall introduce $LD\ $actions on the sigma
model phase spaces $LTG$ and $LTG^{\ast}$ for $G$ and $G^{\ast}$ being dual
Poisson-Lie groups composing a (connected, simply connected) perfect Drinfeld
double $D$, i.e., it admits a global factorization $D=G\bowtie G^{\ast}$.
Under this conditions we have the exact sequences%
\begin{align*}
0  &  \longrightarrow\mathfrak{g}\longrightarrow\mathfrak{d}\longrightarrow
\mathfrak{g}^{\ast}\longrightarrow0\\
0  &  \longrightarrow G\longrightarrow D\longrightarrow G^{\ast}%
\longrightarrow0
\end{align*}
where $\mathfrak{d}=\mathfrak{g}\bowtie\mathfrak{g}^{\ast}$ is the Lie
bialgebra of $D$ supplied with the symmetric invariant no degenerate bilinear
form $(,)_{\mathfrak{d}}$ given by the pairing between $\mathfrak{g}$%
\emph{\ }and\emph{\ }$\mathfrak{g}^{\ast}$, so they are\emph{\ isotropic
subspaces} in relation to it. Identifying $\mathfrak{g}^{\ast}\ $with the Lie
algebra of $G^{\ast}$, we can have the embedding%
\[
L(G\times\mathfrak{g}^{\ast})\overset{\iota_{G}}{\hookrightarrow}LD\times
L\mathfrak{d}^{\ast}~~/~~~(g,\alpha)\longmapsto(g,Ad_{g^{-1}}^{LD\ast}%
\psi(\alpha)+C_{k}(g))
\]
and define the map $\mu:$ $L(G\times\mathfrak{g}^{\ast})\longrightarrow
L\mathfrak{d}_{Aff}^{\ast}$ given by the diagram%
\[
\begin{diagram}[h=1.9em]
L(G \times \mathfrak{g}^{\ast} ) &&\rTo^{i_G}&& LD \times L\mathfrak{d}^{\ast} &\\
&&&&&\\
&&\rdTo_{\mu}&&\dTo_{s}&\\
&&&&\\
&&&& L\mathfrak{d}^{\ast}_{Aff}&\\ \end{diagram}
\]
where $s(l,\xi)=\xi$ is a Poisson map (in fact, it is the source map for the
symplectic groupoid $H=LD\rhd L\mathfrak{d}^{\ast}\rightrightarrows
L\mathfrak{d}_{Aff}^{\ast}$). From the isotropy of $\mathfrak{g}$ with respect
to $(,)_{\mathfrak{d}},$ it can be easily seen that $\iota_{G}^{\ast}%
\omega_{\Gamma}^{R}=\omega_{o}^{LG}$ where $\omega_{o}^{LG}$ is the standard
symplectic structure on $LT^{\ast}G\sim L(G\times\mathfrak{g}^{\ast})$
trivialized by left translations. So $\mu$ is a \emph{Poisson map} for this
symplectic structure on $L(G\times\mathfrak{g}^{\ast}),$ and an analogous
construction can be repeated on the dual group giving the \emph{Poisson map}
$\tilde{\mu}:$ $L(G^{\ast}\times\mathfrak{g})\longrightarrow L\mathfrak{d}%
_{Aff}^{\ast}$. This maps can be regarded as giving symplectic realizations of
$L\mathfrak{d}_{Aff}^{\ast}$ and, as we shall see in the next sections, they
give (non equivariant) momentum maps for $LD$ actions on $L(G\times
\mathfrak{g}^{\ast})\ $and $L(G^{\ast}\times\mathfrak{g})$. For simplicity,
the centrally extended loop group $LD^{\wedge}$ is assumed to exist, so
$(\mu,1)$ and $(\tilde{\mu},1)$ give the usual equivariant momentum maps for
the corresponding $LD^{\wedge}$ actions, however we remark that all the
following constructions can be performed also without using $LD^{\wedge}$ at
all, just replacing $L\mathfrak{d}_{\Gamma}^{\ast}$ by $L\mathfrak{d}%
_{Aff}^{\ast}$.

\subsection{Hamiltonian $LD^{\wedge}$ action on the $G$-sigma model phase
space}

In this section we introduce a $LD^{\wedge}$ symmetry on the sigma model with
target $G$. One of the most striking features of the double Lie groups and Lie
bialgebras is the existence of reciprocal actions between the factors $G$ and
$G^{\ast}$ called \emph{dressing actions} \cite{STS,Lu-We},\cite{Lu-We}.
Writing every element $l\in D$ as $l=g\tilde{h}$, with $g\in G$ and $\tilde
{h}\in G^{\ast}$, the product $\tilde{h}g$ in $D$ can be written as $\tilde
{h}g=g^{\tilde{h}}\tilde{h}^{g}$, with $g^{\tilde{h}}\in G$ and $\tilde{h}%
^{g}\in G^{\ast}$. The dressing action of $G^{\ast}$ on $G$ is then defined
as
\begin{equation}
\mathsf{Dr}:G^{\ast}\times G\longrightarrow G\qquad/\qquad\mathsf{Dr}\left(
\tilde{h},g\right)  =g^{\tilde{h}} \label{dress-G*}%
\end{equation}
The infinitesimal generator of this action in the point $g\in G$ is, for
$\xi\in\mathfrak{g}^{\ast}$,
\[
\xi\longrightarrow\mathsf{dr}\left(  \xi\right)  _{g}=-\left.  \dfrac{d}%
{dt}\mathsf{Dr}\left(  e^{t\xi},g\right)  \right\vert _{t=0}%
\]
such that, for $\eta\in\mathfrak{g}^{\ast}$, we have $\left[  \mathsf{dr}%
\left(  \xi\right)  _{g},\mathsf{dr}\left(  \eta\right)  _{g}\right]
=\mathsf{dr}\left(  \left[  \xi,\eta\right]  _{\mathfrak{g}^{\ast}}\right)
_{g}$. It satisfy the relation
\begin{equation}
Ad_{g^{-1}}^{D}\xi=-\left(  L_{g^{-1}}\right)  _{\ast}\mathsf{dr}\left(
\xi\right)  _{g}+Ad_{g}^{\ast}\xi\label{Ad=dress+}%
\end{equation}
where $Ad_{g^{-1}}^{D}\xi\in\mathfrak{g}\bowtie\mathfrak{g}^{\ast}$ is the
adjoint action of $D$. Then, we can write $\mathsf{dr}\left(  \xi\right)
_{g}=-\left(  L_{g}\right)  _{\ast}\Pi_{\mathfrak{g}}Ad_{g^{-1}}^{D}\xi$, with
$\Pi_{\mathfrak{g}}:\mathfrak{g}\oplus\mathfrak{g}^{\ast}\rightarrow
\mathfrak{g}$ being the projector.

From this dressing action we build up a symplectic action of the double
$LD^{\wedge}$ on $T^{\ast}LG$ whose momentum map furnish the arrow $\mu$ in
diagram $\left(  \ref{T-duality-1}\right)  $. First, we introduce the map
$\mathsf{d}^{LG}:LD\times LG\longrightarrow LG$\ defined as
\begin{equation}
\mathsf{d}^{LG}\left(  a\tilde{b},g\right)  =ag^{\tilde{b}} \label{dr-LG-1}%
\end{equation}
for $a,g\in LG$ and $\tilde{b}\in LG^{\ast}$, which is a left action. It can
be lifted to the left trivialization of $LT^{\ast}G$, namely $L\left(
G\times\mathfrak{g}^{\ast}\right)  $, and then promoted to an action of the
centrally extended double $LD^{\wedge}\simeq LD\times\mathbb{T}^{1}$, as
explained in the following proposition.

\begin{description}
\item \textbf{Proposition:} \textit{The map} $\mathsf{\hat{d}}:LD^{\wedge
}\times L\left(  G\times\mathfrak{g}^{\ast}\right)  \longrightarrow L\left(
G\times\mathfrak{g}^{\ast}\right)  $,
\begin{equation}
\mathsf{\hat{d}}\left(  a\tilde{b},\left(  g,\eta\right)  \right)  =\left(
ag^{\tilde{b}},Ad_{\tilde{b}^{g}}^{LD}\eta+k\left(  \tilde{b}^{g}\right)
^{\prime}\left(  \tilde{b}^{g}\right)  ^{-1}\right)  \label{dr-LG-E-1}%
\end{equation}
\textit{is a left symplectic action, with }$\widehat{Ad}^{LD}$%
\textit{-equivariant momentum mapping}
\begin{equation}
\mu\left(  g,\eta\right)  =\widehat{Ad}_{g^{-1}}^{LD\ast}\left(  \psi\left(
\eta\right)  ,1\right)  ~~~. \label{dr-LG-E-2}%
\end{equation}

\end{description}

\textbf{Proof: }In order\textbf{\ }to obtain $\mathsf{\hat{d}}$, we lift the
action $\mathsf{d}^{LG}$, eq. $\left(  \ref{dr-LG-1}\right)  $, to $T^{\ast}LG
$. In doing so, we consider the associated map $\mathsf{d}_{a\tilde{b}}%
^{LG}:LG\longrightarrow LG~$such that$~\mathsf{d}_{a\tilde{b}}^{LG}\left(
g\right)  \equiv\mathsf{d}^{LG}\left(  a\tilde{b},g\right)  $. Its
differential is $\left(  \mathsf{d}_{a\tilde{b}}^{LG}\right)  _{\ast
g}v=\left(  L_{ag^{\tilde{b}}}\right)  _{\ast}Ad_{\left(  \tilde{b}%
^{g}\right)  ^{-1}}^{LG^{\ast}\ast}\left(  g^{-1}v_{g}\right)  $ for any
tangent vector $v\in T_{g}G$, from where we obtain the pullback on a 1-forms
$\alpha$ in the transformed point $ag^{\tilde{b}}$
\[
\left(  \mathsf{d}_{a\tilde{b}}^{LG}\right)  ^{\ast}\left(  ag^{\tilde{b}%
},\alpha\right)  =\left(  g,\left(  L_{g^{-1}}\right)  ^{\ast}Ad_{\left(
\tilde{b}^{g}\right)  ^{-1}}^{LG^{\ast}}\left(  L_{ag^{\tilde{b}}}\right)
^{\ast}\alpha\right)
\]
In body coordinates $LG\times L\mathfrak{g}^{\ast}$, and after a change of
variables, we get $\mathsf{d}:LD\times L\left(  G\times\mathfrak{g}^{\ast
}\right)  \longrightarrow L\left(  G\times\mathfrak{g}^{\ast}\right)  $
\begin{equation}
\mathsf{d}\left(  a\tilde{b},\left(  g,\eta\right)  \right)  =\left(
ag^{\tilde{b}},Ad_{\tilde{b}^{g}}^{LD}\eta\right)  \label{dr-LG-2}%
\end{equation}
which is a well defined action. The momentum map $\mu_{o}:L\left(
G\times\mathfrak{g}^{\ast}\right)  \longrightarrow L\mathfrak{d}^{\ast}$
associated to this action is easily calculated from the infinitesimal
generator of the action $\left(  \ref{dr-LG-1}\right)  $
\[
\left(  \mathsf{d}_{g}^{LG}\right)  _{\ast e}\left(  X,\xi\right)  =\left(
R_{g}\right)  _{\ast}X-\mathsf{dr}\left(  \xi\right)  _{g}%
\]
for any $\left(  X,\xi\right)  \in L\mathfrak{d}=L\left(  \mathfrak{g}%
\bowtie\mathfrak{g}^{\ast}\right)  $, with $\mathsf{d}_{g}^{LG}\left(
a\tilde{b}\right)  \equiv\mathsf{d}^{LG}\left(  a\tilde{b},g\right)  $. Using
the expression $\left(  \ref{Ad=dress+}\right)  $, and by the identification
$\psi:L\mathfrak{d}\rightarrow L\mathfrak{d}^{\ast}$ provided by the bilinear
form $\left(  ,\right)  _{L\mathfrak{d}}$, we have
\[
\mu_{o}\left(  g,\eta\right)  =\psi\left(  Ad_{g}^{LD}\eta\right)
\]
which obviously is $Ad^{LD}$-equivariant since it is associated to the lift to
the cotangent bundle of an action on $LG$. \smallskip

The action $\mathsf{d}$ is promoted to an action of the centrally extended
double $LD^{\wedge}\simeq LD\times\mathbb{T}^{1}$, $\mathsf{\hat{d}}^{L\left(
G\times\mathfrak{g}^{\ast}\right)  }:LD^{\wedge}\times L\left(  G\times
\mathfrak{g}^{\ast}\right)  \longrightarrow L\left(  G\times\mathfrak{g}%
^{\ast}\right)  $ by the definition
\[
\mathsf{\hat{d}}\left(  \left(  a\tilde{b},\theta\right)  ,\left(
g,\eta\right)  \right)  =\left(  ag^{\tilde{b}},Ad_{\tilde{b}^{g}}^{LD}%
\eta+k\left(  \tilde{b}^{g}\right)  ^{\prime}\left(  \tilde{b}^{g}\right)
^{-1}\right)
\]
where the element $\theta\in\mathbb{T}^{1}$\ acts trivially, so it descends to
an $LD$ action and by this reason it is omitted in $\left(  \ref{dr-LG-E-1}%
\right)  $. It is worth remarking that $\mathsf{\hat{d}}$ is no longer a lift
of a transformation on $LG$.

The infinitesimal action of some $\left(  X,\xi\right)  \in L\left(
\mathfrak{g}\oplus\mathfrak{g}^{\ast}\right)  $ at the point $\left(
g,\eta\right)  \in L\left(  G\times\mathfrak{g}^{\ast}\right)  $ is
straightforwardly computed, giving%
\begin{equation}
\left(  \mathsf{\hat{d}}_{\left(  g,\eta\right)  }\right)  _{\ast e}\left(
X,\xi\right)  =\left(  \left(  R_{g}\right)  _{\ast}X-\mathsf{dr}\left(
\xi\right)  _{g},\left[  Ad_{g}^{LG\ast}\xi,\eta\right]  _{L\mathfrak{g}%
^{\ast}}+Ad_{g}^{LG\ast}ad_{g^{\prime}g^{-1}}^{L\mathfrak{d}\ast}\xi
+Ad_{g}^{LG\ast}\xi^{\prime}\right)  ~~~~, \label{d-inf-gen}%
\end{equation}
and from this expression we calculate the momentum map $\mu:LG\times
L\mathfrak{g}^{\ast}\longrightarrow L\left(  \mathfrak{g}\bowtie
\mathfrak{g}^{\ast}\right)  _{\Gamma}^{\ast}$ using the canonical symplectic
structure $\omega_{o}$ on $L\left(  G\times\mathfrak{g}^{\ast}\right)  $,
obtaining
\[
\mu\left(  g,\eta\right)  =\left(  Ad_{g^{-1}}^{LD\ast}\psi\left(
\eta\right)  +k\psi\left(  g^{\prime}g^{-1}\right)  ,1\right)  =\widehat
{Ad}_{g^{-1}}^{LD\ast}\left(  \psi\left(  \eta\right)  ,1\right)
\]
that satisfy the $\widehat{Ad}$-equivariance relation%
\[
\mu\left(  \mathsf{\hat{d}}\left(  a\tilde{b},\left(  g,\eta\right)  \right)
\right)  =\widehat{Ad}_{\left(  a\tilde{b}\right)  ^{-1}}^{LD\ast}\mu\left(
g,\eta\right)
\]
Obviously, since $\left(  \mathsf{\hat{d}}_{\left(  g,\eta\right)  }\right)
_{\ast}\left(  X,\xi\right)  $ are Hamiltonian for all $\left(  X,\xi\right)
\in L\mathfrak{d}_{\Gamma}$, the Lie derivative $\mathbf{L}_{\left(
\mathsf{\hat{d}}_{\left(  g,\eta\right)  }\right)  _{\ast}\left(
X,\xi\right)  }\omega_{o}=0$ meaning that $\mathsf{\hat{d}}^{L\left(
G\times\mathfrak{g}^{\ast}\right)  }$ leaves the canonical symplectic form
invariant.$\blacksquare$

Now, some remarks are in order. First, note that $\mathsf{\hat{d}}$ is not a
\emph{free }action, the subgroup $G^{\ast}$ leaves invariant the point
$\left(  e,0\right)  $. Then, observe that if the central extension of the
loop group does not exist the above proposition still defines a hamiltonian
$LD$ action. We can proceed in an analogous manner by using the affine Poisson
structure and affine coadjoint $LD$ actions. Finally, $\left(  T^{\ast
}LG,\omega_{o},\mathsf{\hat{d}},\mu\right)  $ is a \emph{Hamiltonian}
$LD^{\wedge}$\emph{-space}, with $\mu$ equivariant and the image through $\mu$
of $T^{\ast}LG$ is a union of coadjoint orbits in $L\mathfrak{d}_{\Gamma
}^{\ast}$.

\bigskip

\subsection{Factorizing $\hat{\Phi}:\Omega D\rightarrow\mathcal{O}\left(
0,1\right)  $ through $LT^{\ast}G$}

\bigskip In this section we shall show that $\hat{\Phi}:\Omega D\rightarrow
\mathcal{O}\subset L\mathfrak{d}_{\Gamma}^{\ast}$ can be factorized through
$\mu:LT^{\ast}G\rightarrow L\mathfrak{d}_{\Gamma}^{\ast}$ on the pure central
extension coadjoint orbit, composing a three vertices commutative diagram like
the left triangle of $\left(  \ref{T-duality-1}\right)  $.

By definition $\mathcal{O}\equiv\mathcal{O}_{\left(  0,1\right)  }=\left\{
\widehat{Ad}_{\left(  a\tilde{b}\right)  ^{-1}}^{LD\ast}\left(  0,1\right)
~/~a\tilde{b}\in LD\right\}  $ and any point $\left(  g,\eta\right)  \in
\mu^{-1}\left(  \mathcal{O}\right)  $ is, due to the equivariance of $\mu$, of
the form $\left(  g,\eta\right)  =\mathsf{\hat{d}}\left(  a\tilde{b},\left(
e,0\right)  \right)  $ for some $a\tilde{b}\in LD$ implying that $\mu
^{-1}\left(  \mathcal{O}\right)  $ is just the orbit of $LD$ through the point
$\left(  e,0\right)  \in L\left(  G\times\mathfrak{g}^{\ast}\right)  $. In
terms of the orbit map $\mathsf{\hat{d}}_{\left(  e,0\right)  }%
:LD\longrightarrow L\left(  G\times\mathfrak{g}^{\ast}\right)  $,
\[
\mathsf{\hat{d}}_{\left(  e,0\right)  }\left(  a\tilde{b}\right)
=\mathsf{\hat{d}}\left(  a\tilde{b},\left(  e,0\right)  \right)
=(a,k\tilde{b}^{\prime}\tilde{b}^{-1})
\]
we write
\[
\mu^{-1}\left(  \mathcal{O}\right)  =\operatorname{Im}\mathsf{\hat{d}%
}_{\left(  e,0\right)  }%
\]
Hence, the tangent space of this $LD$-orbit is spanned by the infinitesimal
generators of the action $\mathsf{\hat{d}}$, given in eq. $\left(
\ref{d-inf-gen}\right)  $, for every $\left(  g,\eta\right)  =(a,k\tilde
{b}^{\prime}\tilde{b}^{-1})\in\mu^{-1}\left(  \mathcal{O}\right)  $, and it
can be split in the direct sum $\mathsf{\hat{d}}\left(  \Lambda_{\ast
}L\mathfrak{d}\right)  \oplus\mathsf{\hat{d}}\left(  Ad_{a\tilde{b}}%
^{LD}\mathfrak{d}\right)  $. In fact, for any tangent vector $\left(
V,\xi\right)  $ to that point there exist some $\left[  X\right]  \in
\Lambda_{\ast}L\mathfrak{d}$ and $X_{o}\in\mathfrak{d}$ such that
\begin{equation}
\left(  V,\xi\right)  _{\left(  g,\eta\right)  }=\left(  \mathsf{\hat{d}%
}_{\left(  g,\eta\right)  }\right)  _{\ast e}\left(  \left[  X\right]
\right)  +\left(  \mathsf{\hat{d}}_{\left(  g,\eta\right)  }\right)  _{\ast
e}\left(  Ad_{a\tilde{b}}^{LD}X_{o}\right)  \label{split-tang-space}%
\end{equation}
Observe that $\left(  \mathsf{\hat{d}}_{\left(  e,0\right)  }\right)  ^{\ast
}\omega_{o}=\gamma$, eq. $\left(  \ref{gamma}\right)  $ and, beside to the
fact that $\left.  \omega_{o}\right\vert _{\mu^{-1}\left(  \mathcal{O}\right)
}=\mu^{\ast}\omega_{KK}$ , it amounts to a presymplectic submersion $\mu
\circ\mathsf{\hat{d}}_{\left(  e,0\right)  }:\left(  LD,\gamma\right)
\longrightarrow\left(  \mathcal{O},\omega_{KK}\right)  $.

\begin{description}
\item \bigskip\textbf{Theorem: }\textit{Let }$\left.  \omega_{o}\right\vert
_{\mu^{-1}\left(  \mathcal{O}\right)  }$ \textit{the restriction of }%
$\omega_{o}$ \textit{to} $\mu$ $^{-1}\left(  \mathcal{O}\right)  $.
\textit{Then}, \textit{its null distribution is spanned by the infinitesimal
generator of subgroup }$Ad_{a\tilde{b}}^{LD}D$ \textit{with leaf through
}$(a,k\tilde{b}^{\prime}\tilde{b}^{-1})\in\mu^{-1}\left(  \mathcal{O}\right)
$ \textit{being } $\mu^{-1}(\widehat{Ad}_{\left(  a\tilde{b}\right)  ^{-1}%
}^{LD\ast}\left(  0,1\right)  )=\mathsf{\hat{d}}^{L\left(  G\times
\mathfrak{g}^{\ast}\right)  }\left(  \left[  a\tilde{b}\right]  ,\left(
G,0\right)  \right)  ,$ \textit{so }$\mu^{-1}\left(  \mathcal{O}\right)
\longrightarrow\mathcal{O}$ \textit{is a fibration with }$\dim\mathfrak{g}%
$\textit{-dimensional} \textit{fibers. Moreover, their symplectic leaves are}
\textit{the orbits of }$\Omega D$ \textit{by the action} $\mathsf{\hat{d}}$.
\end{description}

\textbf{Proof: }The isotropy group of a point $\widehat{Ad}_{\left(
a\tilde{b}\right)  ^{-1}}^{LD\ast}\left(  0,1\right)  \in\mathcal{O}$ is
$Ad_{a\tilde{b}}^{LD}D$, and its infinitesimal action pulled-back by $\mu$
gives rise to the null distribution of $\left.  \omega_{o}\right\vert
_{\mu\left(  \mathcal{O}\left(  0,1\right)  \right)  }$. So, we have the null
foliation with leaf through $\mathsf{\hat{d}}\left(  a\tilde{b},\left(
e,0\right)  \right)  $ being the orbits of the subgroup $Ad_{a\tilde{b}}%
^{LD}D$ and of dimension $\dim\mathfrak{g}$, as it can be seen from the
relation
\[
\mathsf{\hat{d}}^{L\left(  G\times\mathfrak{g}^{\ast}\right)  }\left(
(a\tilde{b})l_{o}(a\tilde{b})^{-1},(a,k\tilde{b}^{\prime}\tilde{b}%
^{-1})\right)  =\mathsf{\hat{d}}^{L\left(  G\times\mathfrak{g}^{\ast}\right)
}\left(  (a\tilde{b})l_{o},(e,0)\right)  =\mathsf{\hat{d}}^{L\left(
G\times\mathfrak{g}^{\ast}\right)  }\left(  (a\tilde{b}),(g_{o},0)\right)
\]
for all $l_{o}=g_{o}\tilde{h}_{o}\in D$ or, infinitesimally, since for
$\left(  g,\eta\right)  =\mathsf{\hat{d}}\left(  a\tilde{b},\left(
e,0\right)  \right)  $,%
\[
\left(  \mathsf{\hat{d}}_{\left(  g,\eta\right)  }\right)  _{\ast e}\left(
Ad_{a\tilde{b}}^{LD}X_{o}\right)  =\left(  \mathsf{\hat{d}}_{\mathsf{\hat{d}%
}\left(  a\tilde{b},\left(  e,0\right)  \right)  }\right)  _{\ast e}\left(
Ad_{a\tilde{b}}^{LD}X_{o}\right)  =\left(  \mathsf{\hat{d}}_{a\tilde{b}%
}\right)  _{\ast\left(  e,0\right)  }\left(  \mathsf{\hat{d}}_{\left(
e,0\right)  }\right)  _{\ast e}X_{o}%
\]
and from $\left(  \ref{d-inf-gen}\right)  \ \left(  \mathsf{\hat{d}}_{\left(
e,0\right)  }\right)  _{\ast e}X_{o}=\left(  \Pi_{\mathfrak{g}}(X_{o}%
),0\right)  $ for all $X_{o}\in\mathfrak{d}$.

The complementary distribution constitutes then the symplectic foliation of
$\left.  \omega_{o}\right\vert _{\mu^{-1}\left(  \mathcal{O}\right)  }$, and
it is spanned by the second term in the direct sum $\left(
\ref{split-tang-space}\right)  $ with leaves being the orbits of $\Omega D$ in
$\mu^{-1}\left(  \mathcal{O}\right)  $.

All this can be obtained from a more general result contained in a theorem by
Kazhdan, Kostant and Sternberg \cite{KKS} (see Theorem 26.2 in ref.
\cite{Guill-Sten}).$\blacksquare$

A further consequence is that any point $\left(  g,\eta\right)  \in$ $\mu
^{-1}\left(  \mathcal{O}\right)  $ can be characterized by a pair $\left(
\left[  a\tilde{b}\right]  ,g_{o}\right)  \in\Omega D\times G$ such that
$\left(  g,\eta\right)  =\mathsf{\hat{d}}^{L\left(  G\times\mathfrak{g}^{\ast
}\right)  }\left(  \left[  a\tilde{b}\right]  ,\left(  g_{o},0\right)
\right)  $. \bigskip Thinking of the composition
\[
\mu^{-1}\left(  \mathcal{O}\right)  \overset{\mu}{\longrightarrow}%
\mathcal{O}\overset{\hat{\Phi}^{-1}}{\longrightarrow}\Omega D
\]
as a fibration, the direct sum $\left(  \ref{split-tang-space}\right)  $
defines a connection with the horizontal subspace being the orbits $S(g_{o})$
of $\Omega D$ through the point $\left(  g_{o},0\right)  $, for each $g_{o}\in
G$. Then, we may consider a section $\zeta:\Omega D\rightarrow\mu^{-1}\left(
\mathcal{O}\right)  $ with equivariance property%
\[
\zeta\left(  l\cdot\left[  m\right]  \right)  =\mathsf{\hat{d}}\left(
l,\zeta\left(  \left[  m\right]  \right)  \right)  ~~~,
\]
to obtain a factorization of $\hat{\Phi}$ through $T^{\ast}LG$,%

\begin{equation}
\begin{diagram}[h=1.9em] & & && (L{\mathfrak{d}}^{\ast};\;\{,\}_{KK})&\\ &&\ruTo^{\mu} && & \\ (T^{\ast }LG;\omega _{o}) & &&&\uTo^{\hat{\Phi} }&\\ &&\luTo_{\phi} &&&\\ &&&&(\Omega D;\omega_{\Omega D} )&\\ \end{diagram} \label{diag-left}%
\end{equation}
with arrows being presymplectic maps, reproducing the left triangle of diagram
$\left(  \ref{T-duality-1}\right)  $.

In order to get a symplectic factorization, we define a family $\left\{
\varsigma_{g_{o}}\right\}  _{g_{o}\in G}$ of \emph{horizontal} sections with
each image being a symplectic leaf $S(g_{o})\subset\mu^{-1}\left(
\mathcal{O}\right)  $, the label $g_{o}$ being the point $\left(
g_{o},0\right)  $ in $\mu^{-1}\left(  \mathcal{O}\right)  $ they pass through
\begin{align*}
\varsigma_{g_{o}}  &  :\Omega D\longrightarrow S(g_{o})\subset\mu^{-1}\left(
\mathcal{O}\right) \\
\varsigma_{g_{o}}\left(  \left[  l\right]  \right)   &  =\mathsf{\hat{d}%
}\left(  \left[  l\right]  ,\left(  g_{o},0\right)  \right)
\end{align*}
for $g_{o}\in G$. Indeed, they are horizontal sections if one regards $\left(
\ref{split-tang-space}\right)  $ as defining a connection on the trivial
$G$-bundle structure in $\mu^{-1}\left(  \mathcal{O} \right)  $.

\smallskip

\begin{description}
\item \textbf{Proposition: }\textit{Let }$\varsigma_{g_{o}}:\left(  \Omega
D,\omega_{\Omega D}\right)  \longrightarrow\left(  S(g_{o}),\omega
_{o}|_{S(g_{o})}\right)  $\textit{\ is} \textit{a symplectic map}, \textit{for
any }$g_{o}\in G$\textit{.}
\end{description}

\textbf{Proof: }We have to show that $\omega_{\Omega D}=\varsigma_{g_{o}%
}^{\ast}\omega_{o}|_{S(g_{o})}$. Let $v\in T_{l}LD$, and $\left[  v\right]
=\Lambda_{\ast l}v\in T_{\left[  l\right]  }\Omega D$ a tangent vector to the
point $\left[  l\right]  \in\Omega D$. Then%
\[
\left(  \varsigma_{g_{o}}\right)  _{\ast\left[  l\right]  }\left(  \left[
v\right]  _{b}\right)  =\left(  \mathsf{\hat{d}}_{\varsigma_{g_{o}}\left(
\left[  l\right]  \right)  }\right)  _{\ast}\left(  \left[  v\right]  \left[
l\right]  ^{-1}\right)
\]
where $\mathsf{\hat{d}}_{\varsigma_{g_{o}}\left(  \left[  l\right]  \right)
}:LD\longrightarrow T^{\ast}LG$ is the induced map by the action
$\mathsf{\hat{d}}$ describing the orbit of $LD$ through $\left(
g_{o},0\right)  \in L\left(  G\times\mathfrak{g}^{\ast}\right)  $. Using the
equivariant momentum map $\mu$ and having in mind that the stabilizer of the
point $\widehat{Ad}_{\left[  a\tilde{b}\right]  ^{-1}}^{LD\ast}\left(
0,1\right)  $ is $Ad_{a\tilde{b}}^{LD\ast}D$, we conclude that
\begin{align*}
\left\langle \left.  \omega_{o}\right\vert _{\mu^{-1}\left(  \mathcal{O}%
\right)  },\left(  \varsigma_{g_{o}}\right)  _{\ast\left[  l\right]  }\left(
\left[  v\right]  \right)  \otimes\left(  \varsigma_{g_{o}}\right)
_{\ast\left[  l\right]  }\left(  \left[  w\right]  \right)  \right\rangle  &
=\left\langle Ad_{l^{-1}}^{LD\ast}\left(  0,1\right)  ,\left[  \left[
v\right]  \left[  l\right]  ^{-1},\left[  w\right]  \left[  l\right]
^{-1}\right]  _{L\mathfrak{d}_{\Gamma}}\right\rangle \\
&  =\Gamma_{k}\left(  \left[  l\right]  ^{-1}\left[  v\right]  ,\left[
l\right]  ^{-1}\left[  w\right]  \right)
\end{align*}
showing that $\left(  \varsigma_{g_{o}}\right)  ^{\ast}\left.  \omega
_{o}\right\vert _{S_{R}(g_{o})}=\omega_{\Omega D}$. $\blacksquare$

Then, the diagram $\left(  \ref{diag-left}\right)  $ can be refined to the
following commutative diagram with arrows being symplectic maps:%

\begin{equation}
\begin{diagram}[h=1.9em] &&& &(\mathcal{O};\omega_{KK})&\\ &&\ruTo^{\mu} && &\\ \left({\mu}^{-1}\left(\mathcal{O}\right);\omega_{o}\right) &&&&\uTo^{\hat{\Phi}}&\\ &&\luTo_{\varsigma_{g_{o}}}&&&\\ &&&&\left(\Omega D;\omega_{\Omega D} \right)&\\ \end{diagram} \label{diag-lef-refined}%
\end{equation}
for each $g_{o}\in G$.

\subsection{Hamiltonian $LD^{\wedge}$ action on the $G^{\ast}$-sigma model
phase space}

Because of the symmetric role played by $G$ and $G^{\ast}$in the double
$D,$all the results obtained above can be straightforwardly dualized just
interchanging their roles. In spite of this general principle, a few details
and notation are in order.

\smallskip Let us consider $D$ with the opposite factorization, denoted as
$D\rightarrow D^{T}=G^{\ast}\bowtie G$, so that every element is now written
as $\tilde{h}g$ with $\tilde{h}\in G^{\ast}$ and $g\in G$. Then, for $g\in G $
and $\tilde{h}\in G^{\ast}$ there exist $\tilde{h}_{g}\in G^{\ast}$ and
$g_{\tilde{h}}\in G$ such that $g\tilde{h}=\tilde{h}_{g}g_{\tilde{h}}$. This
factorization relates with the opposite one by $\tilde{b}_{a}=\left(  \left(
\tilde{b}^{-1}\right)  ^{a^{-1}}\right)  ^{-1}$ and $a_{\tilde{b}}=\left(
\left(  a^{-1}\right)  ^{\tilde{b}^{-1}}\right)  ^{-1}$. The dressing action
$\widetilde{\mathsf{Dr}}:G\times G^{\ast}\longrightarrow G^{\ast}$ is
$\widetilde{\mathsf{Dr}}\left(  g,\tilde{h}\right)  =\tilde{h}_{g}$ and, by
composing it with the right action of $LG^{\ast}$ on itself, we get the action
$\mathsf{b}^{LG^{\ast}}:LD\times LG^{\ast}\longrightarrow LG^{\ast}\ $defined
as $\mathsf{b}^{LG}\left(  \tilde{b}a,\tilde{h}\right)  =\tilde{b}\tilde
{h}_{a}$\textit{\ }with $a\in LG$ and $\tilde{h},\tilde{b}\in LG^{\ast}$.

For a sigma model on the target $G^{\ast}$, the phase space is the symplectic
manifold $\left(  T^{\ast}LG^{\ast},\tilde{\omega}_{o}\right)  $. Then, we
lift $\mathsf{b}^{LG^{\ast}}$ to the left trivialization of $T^{\ast}LG^{\ast
}\cong$ $L\left(  G^{\ast}\times\mathfrak{g}\right)  $, and promote it to an
extended symmetry $\mathsf{\hat{b}}:LD^{\wedge}\times L\left(  G^{\ast}%
\times\mathfrak{g}\right)  \longrightarrow L\left(  G\times\mathfrak{g}^{\ast
}\right)  $,%
\begin{equation}
\mathsf{\hat{b}}\left(  \tilde{b}a,\left(  \tilde{h},Z\right)  \right)
=\left(  \tilde{b}\tilde{h}_{a},Ad_{a_{\tilde{h}}}^{LD}Z+k\left(  a_{\tilde
{h}}\right)  ^{\prime}\left(  a_{\tilde{h}}\right)  ^{-1}\right)
\label{dr-LG*-E-1}%
\end{equation}
with an $\widehat{Ad}^{LD}$-equivariant momentum map
\begin{equation}
\tilde{\mu}\left(  \tilde{h},Z\right)  =\widehat{Ad}_{\tilde{h}^{-1}}^{LD\ast
}\left(  \psi\left(  Z\right)  ,1\right)  =\left(  \psi\left(  Ad_{\tilde{h}%
}^{LD}Z+k~\tilde{h}^{\prime}\tilde{h}^{-1}\right)  ,1\right)  \,\,
\label{dr-LG*-E-2}%
\end{equation}

In terms of the orbit map $\mathsf{\hat{b}}_{\left(  e,0\right)
}:LD\longrightarrow L\left(  G^{\ast}\times\mathfrak{g}\right)  $ associated
to the action $\mathsf{\hat{b}}$, eq. $\left(  \ref{dr-LG*-E-1}\right)  $,%
\[
\mathsf{\hat{b}}_{\left(  e,0\right)  }\left(  a\tilde{b}\right)
=\mathsf{\hat{b}}\left(  \tilde{b}a,\left(  e,0\right)  \right)  =(\tilde
{b},k~a^{\prime}a^{-1})
\]
and we get the identification
\[
\mu^{-1}\left(  \mathcal{O}\right)  =\operatorname{Im}\mathsf{\hat{b}%
}_{\left(  e,0\right)  }%
\]

Analogously, we define an equivariant map $\tilde{\zeta}:LD\rightarrow
L\left(  G^{\ast}\times\mathfrak{g}\right)  $ such that $\tilde{\zeta}\left(
l\cdot\left[  m\right]  \right)  =\mathsf{\hat{b}}\left(  l,\tilde{\zeta
}\left(  \left[  m\right]  \right)  \right)  ~~$and get a factorization of
$\hat{\Phi}$ through $T^{\ast}LG^{\ast}$ as $\hat{\Phi}=\tilde{\mu}\circ
\tilde{\varsigma}_{\tilde{h}_{o}}$, by symplectic maps, depicted in the diagram%

\begin{equation}
\begin{diagram}[h=1.9em] (\mathcal{O};\omega_{KK})&& &&&\\ &&\luTo^{\tilde{\mu}} & &&\\ \uTo^{\hat{\Phi}}&&&&(\tilde{\mu}^{-1}(\mathcal{O}); \left.\tilde{\omega}_o\right\vert_{r}) &\\ &&\ruTo_{\tilde{\varsigma}_{{\tilde h}_o}}&&&\\ (\Omega D;\ \omega_{\Omega D })&& && &\\ \end{diagram} \label{diag-right-refined}%
\end{equation}
for each $\tilde{h}_{o}\in G^{\ast}$.

\section{III- Poisson-Lie T-duality}

Gluing together diagrams $\left(  \ref{diag-left}\right)  $ and its mirror
image, we recover the commutative four vertex diagram $\left(
\ref{T-duality-1}\right)  $ relating the phase spaces $LT^{\ast}G$ and
$LT^{\ast}G^{\ast}$, belonging to $\sigma$-models with dual targets $G$ and
$G^{\ast}$, through vertex $(\Omega D,$ $\omega_{\Omega D})$ and
$(L\mathfrak{d}_{\Gamma}^{\ast},\left\{  ,\right\}  _{KK})$ and with arrows
being $LD^{\wedge}$-equivariant (pre)symplectic maps.

Actually, the relation holds provided there exist a non trivial intersection
region of the images of the momentum maps $\mu$ and $\tilde{\mu}$ in $\left(
L\mathfrak{d}_{\Gamma}\right)  ^{\ast}$, that means, if there exist a set of
points $\left(  g,\eta\right)  \in L\left(  G\times\mathfrak{g}^{\ast}\right)
$ and $\left(  \tilde{h},Z\right)  \in L\left(  G^{\ast}\times\mathfrak{g}%
\right)  $ satisfying (see eqs. $\left(  \ref{dr-LG-E-2}\right)  $ and
$\left(  \ref{dr-LG*-E-2}\right)  )$ the identity%
\[
\left(  \psi\left(  \eta\right)  ,1\right)  =\widehat{Ad}_{\tilde{h}^{-1}%
g}^{LD\ast}\left(  \psi\left(  Z\right)  ,1\right)
\]
Because $\mu$ and $\tilde{\mu}$ are equivariant momentum maps, the
intersection region extends to the whole coadjoint orbit of the point
$\mu\left(  g,\eta\right)  =\tilde{\mu}\left(  \tilde{h},Z\right)  $ in
$\left(  L\mathfrak{d}_{\Gamma}\right)  ^{\ast}$, establishing a connection
between $LD^{\wedge}$-orbits in $LT^{\ast}G$ and $LT^{\ast}G^{\ast}$. It can
be seen that this common region coincides with the pure central extension
orbit
\[
\mathcal{O}=\operatorname{Im}\mu\cap\operatorname{Im}\tilde{\mu}%
\]
and it is a isomorphic image of the WZW reduced space, so that we may refine
diagram $\left(  \ref{T-duality-1}\right)  $ to get a connection between the
phase spaces of sigma models on dual targets and WZW model on the associated
Drinfeld double group%

\begin{equation}
\begin{diagram}[h=1.9em] ({\mu}^{-1}(\mathcal{O});\left.\omega_{o}\right\vert _{\mu^{-1}})&&\rTo^{\mu}&&{({\mathcal{O}};\omega_{KK})}&&\lTo^{\tilde{\mu}}&&({\tilde\mu}^{-1}(\mathcal{O});\left.{\tilde\omega}_{o}\right\vert _{\tilde{\mu}^{-1}})\\ &&&&&&&&\\ &\luTo(3,3)_{\varsigma_{g_o}}&&&\dTo^{{\hat{\Phi}}^{-1}}&&&\ruTo(3,3)_{{\tilde\varsigma}_{{\tilde h}_o}}&\\ &&&&&&&&\\ &&&&(\Omega D;\omega_{\Omega D} )&&&&\\ \end{diagram} \label{T-duality-2}%
\end{equation}
\emph{Poisson-Lie T-duality} is then accurately characterized by restricting
this diagram to the symplectic leaves in $\mu^{-1}\left(  \mathcal{O}\right)
$ and $\tilde{\mu}^{-1}\left(  \mathcal{O}\right)  $. In fact, lets us denote
by $S\left(  g_{o}\right)  \subset\mu^{-1}\left(  \mathcal{O}\right)  $ and
$\tilde{S}\left(  \tilde{h}_{o}\right)  \subset\tilde{\mu}^{-1}\left(
\mathcal{O}\right)  $ the symplectic leaves defined by the maps $\varsigma
_{g_{o}}:\Omega D\rightarrow\mu^{-1}\left(  \mathcal{O}\right)  $ and
$\varsigma_{\tilde{h}_{o}}:\Omega D\rightarrow\tilde{\mu}^{-1}\left(
\mathcal{O}\right)  $, respectively, then the composition of arrows%

\begin{equation}
\begin{diagram}[h=1.9em] S(h_{o})&&\rTo^{\mu}&&{\mathcal{O}}&&\lDashto^{\tilde{\mu}}&&\;{\tilde S}({\tilde h}_o)\\ &&&&&&&\;\;\;\;\;\;\;\;\ruTo(4,4)_{{\tilde\varsigma}_{{\tilde h}_o}}\;\;\;\;\;\;\;\;&\\ &&&&\dTo^{{\hat{\Phi}}^{-1}}&&&&\\ &&&&&&&&\\ &&&&\Omega D&&&&\\ \end{diagram} \label{T-duality-3}%
\end{equation}

defines the duality \emph{T-duality} transformation
\begin{align}
\Psi_{\tilde{h}_{o}}  &  :S\left(  g_{o}\right)  \longrightarrow S^{\ast
}\left(  \tilde{h}_{o}\right) \label{T-duality-4}\\
\Psi_{\tilde{h}_{o}}\left(  g,\lambda\right)   &  =\left(  \tilde{b}\left(
\tilde{h}_{o}\right)  _{a},\left(  a_{\tilde{h}_{o}}\right)  ^{\prime}\left(
a_{\tilde{h}_{o}}\right)  ^{-1}\right) \nonumber
\end{align}
where $\left[  a\tilde{b}\right]  \in\Omega D$ is a based loop such that
$\left(  g,\lambda\right)  =\mathsf{\hat{d}}\left(  \left[  a\tilde{b}\right]
,\left(  g_{o},0\right)  \right)  $. This are nothing but the duality
transformations given in \cite{KS-1} and \cite{Sfetsos}. Obviously, as a
composition of symplectic maps, $\Psi_{\tilde{h}_{o}}$ is a canonical
transformation, and a hamiltonian vector field tangent to $S\left(
g_{o}\right)  $ is mapped onto a hamiltonian vector field tangent to
$\tilde{S}\left(  \tilde{h}_{o}\right)  $. Passing through $\Omega D$ by
$\hat{\Phi}^{-1}$, as showed in the diagram, allows to switch $\Omega D$ to
the opposite factorization $\left[  a\tilde{b}\right]  _{B}\rightarrow\left[
\tilde{b}_{a}a_{\tilde{b}}\right]  _{B}$ before to reach $\tilde{S}\left(
\tilde{h}_{o}\right)  $ by applying
\[
\tilde{\varsigma}_{\tilde{h}_{o}}:\left[  a\tilde{b}\right]  _{B}%
\longrightarrow\left[  \tilde{b}_{a}a_{\tilde{b}}\right]  _{B}\longrightarrow
\mathsf{\hat{b}}\left(  \left[  \tilde{b}_{a}a_{\tilde{b}}\right]
_{b},\left(  \tilde{h}_{o},0\right)  \right)
\]
for $\left(  \tilde{h}_{o},0\right)  \in\tilde{\mu}^{-1}\left(  0,1\right)  $.
\bigskip

Observe that diagram $\left(  \ref{T-duality-1}\right)  $ can also be
constructed for an arbitrary bicrossed product $D=G\bowtie M$ with a Lie
algebra $\mathfrak{g}\bowtie\mathfrak{m}$ supplied with a non degenerate,
symmetric, invariant bilinear form, replacing the vertex $LT^{\ast}G$ by
$L(G\times\mathfrak{m}).$ Now this vertex carries a presymplectic structure
defined by the pullback $\iota_{G}^{\ast}\omega_{\Gamma}^{R}$ and
\begin{align*}
&  L(G\times\mathfrak{m})\overset{\iota_{G}}{\hookrightarrow}LD\times
L\mathfrak{d}^{\ast}\\
(g,\alpha)  &  \longmapsto(g,Ad_{g^{-1}}^{D\ast}\psi(\alpha)+C_{k}(g))
\end{align*}
recovering the generalization of PL-T duality introduced in \cite{Majid-Begg}.

\bigskip

\section{IV- Collective Hamiltonians and duality transformations}

Using the \emph{geometrical} or \emph{kinematical} information of the diagram
$\left(  \ref{T-duality-3}\right)  $ we now address to impose the appropriate
dynamics in order it can be mapped through the arrows giving dynamical
T-duality transformations.

To this end , we observe that $\Omega D=\Phi^{-1}\left(  \mathcal{O}\right)  $
and the symplectic leaves in $\mu^{-1}\left(  \mathcal{O}\right)  $,
$\tilde{\mu}^{-1}\left(  \mathcal{O}\right)  $ are replicas of the coadjoint
orbit $\mathcal{O}$ and, because of their equivariance, their tangent bundles
are locally isomorphic to that of $\mathcal{O}$. As $\mathcal{O}$ is in the
vertex linking the three models, it is clear that T-duality transformation
$\left(  \ref{T-duality-3}\right)  $ exist at the level of hamiltonian vector
fields for each hamiltonian vector on $\mathcal{O}$. So, it enough to select a
hamiltonian vector field in $\mathcal{O}$ and symplectic leaves in $\mu
^{-1}\left(  \mathcal{O}\right)  $ and $\tilde{\mu}^{-1}\left(  \mathcal{O}%
\right)  $ to obtain a couple of T-dual related hamiltonian vector fields and,
whenever they exist, a couple T-dual related solution curves belonging to some
kind of dual sigma models.

In terms of hamiltonian functions, once a suitable function $\mathsf{h}%
:\left(  L\mathfrak{d}_{\Gamma}\right)  ^{\ast}\longrightarrow\mathbb{R}$ is
fixed we have the corresponding hamiltonian function on $\mu^{-1}\left(
\mathcal{O}\right)  $ and $\tilde{\mu}^{-1}\left(  \mathcal{O}\right)  $ by
pulling-back it through the momentum maps $\mu$ and $\tilde{\mu}$, so that the
hamiltonian function restricted to $\mu^{-1}\left(  \mathcal{O}\right)  $,
$\tilde{\mu}^{-1}\left(  \mathcal{O}\right)  $ and $\Omega D$ are in the so
called \emph{collective Hamiltonian form }\cite{Guill-Sten}: $\mathsf{h}%
\circ\mu$, $\mathsf{h}\circ\tilde{\mu}$ and $\mathsf{h}\circ\Phi$.

This ensures that the corresponding Hamiltonian vector fields will be tangent
to the $LD$ orbits. Moreover, a Hamiltonian vector field in $\widehat
{Ad}_{l^{-1}}^{LD\ast}\left(  0,1\right)  $ $\in\mathcal{O}$ is of the form
$\widehat{ad}_{\mathcal{L}_{\mathsf{h}}}^{LD\ast}\widehat{Ad}_{l^{-1}}%
^{LD\ast}\left(  0,1\right)  $ where $\mathcal{L}_{\mathsf{h}}:\left(
L\mathfrak{d}_{\Gamma}\right)  ^{\ast}\longrightarrow L\mathfrak{d}_{\Gamma}$
is the corresponding Legendre transformation, and the solution curves are
determined from the solution of the differential equation on $LD$%
\begin{equation}
d_{t}l~l^{-1}=\mathcal{L}_{\mathsf{h}}(\gamma(t))\label{coll-trayect-1}%
\end{equation}
where $\gamma(t)$ is the trajectory of the hamiltonian vector field
corresponding to $\mathsf{h}$. In fact, from the curve $l(t)\subset LD,$ with
$l(t=0)=e$, solution of the equation so that $\gamma(t)=\widehat{Ad}%
_{l(t)}^{LD\ast}\gamma(0)$, the solutions for the collective hamiltonian
vector fields on $T^{\ast}LG,\ \Omega D$ and $T^{\ast}LG^{\ast}$ are
\[
\left\{
\begin{array}
[c]{l}%
\mathsf{\hat{d}}\left(  l(t),\left(  g_{0},\eta_{0}\right)  \right)  \\
\left[  l(t)l_{0}\right]  \\
\mathsf{\hat{b}}\left(  l(t),\left(  \tilde{h}_{0},Z_{0}\right)  \right)
\end{array}
\right.
\]
respectively, with $l(t)\ $given by $\left(  \ref{coll-trayect-1}\right)  $and
for $\left(  g_{o},\eta_{o}\right)  \in\mu^{-1}(\gamma(0)),$ $\left[
l_{0}\right]  \in\hat{\Phi}^{-1}(\gamma(0)),\ $\ $\left(  \tilde{h}_{0}%
,Z_{0}\right)  \in\tilde{\mu}^{-1}(\gamma(0))$.

We see that duality transformations between $T^{\ast}LG$ and $T^{\ast}%
LG^{\ast}$ involve finding the curve $l(t)$ solution to eq. $\left(
\ref{coll-trayect-1}\right)  $ and using the two factorizations of the double
$D=G\bowtie G^{\ast}\sim G^{\ast}$ $\bowtie G$. The generating functional on
the dualizable subspaces is given in terms of the potentials $\vartheta_{o}%
\ $and $\tilde{\vartheta}_{o}$ of the symplectic forms on the dual
phase-spaces by \cite{Alvarez-npb}%
\begin{align*}
d_{V}F[g,\tilde{g}] &  =\left.  \vartheta_{o}-\tilde{\vartheta}_{o}\right\vert
_{S_{R}\times\tilde{S}_{R}}\\
&  =\left\langle g^{-1}dg,\tilde{h}%
\acute{}%
\tilde{h}^{-1}\right\rangle -\left\langle \tilde{g}^{-1}d\tilde{g},h%
\acute{}%
h^{-1}\right\rangle \\
&  =-\int_{S^{1}}l^{\ast}(\iota_{V}\ \omega^{STS})
\end{align*}
where $V$ is a vector field along the loop $l=g\tilde{h}=\tilde{g}h$ in $D$
and $\omega^{STS}$ is the symplectic form on the double $D$ \cite{STS},
\cite{Alek-Malkin}. This leads to the well known generating functional formula
\cite{KS-1} for PLT-duality.

Also note that, in order to have a non trivial duality, restriction to the
common sector in $\left(  L\mathfrak{d}_{\Gamma}\right)  ^{\ast}$ where all
the moment maps intersect is required, i.e., to the coadjoint orbit
$\mathcal{O}$. This is why the study of the pre-images $\mu^{-1}\left(
\mathcal{O}\right)  $ and $\tilde{\mu}^{-1}\left(  \mathcal{O}\right)  $ of
the last section becomes relevant. So, from now on, we shall refer to this
pre-images as \emph{dualizable} or \emph{admissible}{\ subspaces}.

Now, a couple of remarks are in order. First, note that an analogous diagram
to $\left(  \ref{T-duality-1}\right)  $ can be constructed by replacing one of
the phase spaces by any $LD$\symbol{94}-hamiltonian space. The same statements
will hold for collective hamiltonian dynamics and so we can construct the
corresponding duality transformations. This will lead us, as special cases, to
\emph{Buscher
\'{}%
s duality} introduced in \cite{KS-1} and to duality between different
factorizations of the Drinfeld double bialgebra $\mathfrak{d}=m+m^{\ast}$,
some giving the so called PLT-plurality \cite{Plurality}..

We also like to remark, before passing to the next section, that even when the
central extension of the loop group $LD$ does not exist, the same diagrams can
be constructed and all the statements about collective dynamics (and so all
about T-dualtity and duality transformations) still hold after replacement of
the dual of centrally extended loop algebra $\left(  L\mathfrak{d}_{\Gamma
}\right)  ^{\ast}$ by $L\mathfrak{d}_{Aff}^{\ast}$ with the affine Poisson
structure defined by the cocycle $\Gamma$ and the affine coadjoint action
defined in section I.

In the following subsections, we shall study the dynamics of collective
Hamiltonians and the corresponding lagrangian formulation for the T-dual WZW
and sigma models$.$

\subsection{Hamiltonian and Lagrangian WZW model}

The WZW-model reduced space $\Omega D$ is mapped into the coadjoint orbit
$\mathcal{O}$ by the momentum map $\hat{\Phi}:\Omega D\longrightarrow
\mathcal{O}$ associated to the residual left invariance $\left(
\ref{res-l-inv}\right)  $. So, let us consider a Hamiltonian $H_{_{WZW}%
}\left(  g,\eta\right)  $ for the chiral WZW model which, when restricted to
the reduced space $\Omega D=\left(  \hat{J}^{R}\right)  ^{-1}\left(
0,1\right)  /D$, it is in collective form
\[
\left.  H_{_{WZW}}\left(  l,\varphi\right)  \right\vert _{\Omega D}%
=\mathsf{h}\circ\hat{\Phi}\left(  \left[  l\right]  \right)
\]
for some function $\mathsf{h}:\left(  L\mathfrak{d}_{\Gamma}\right)  ^{\ast
}\rightarrow\mathbb{R}$. We shall consider a quadratic Hamiltonian which,
having in mind that $\hat{J}^{R}\left(  l,\varphi\right)  =\left(
k\psi\left(  l^{-1}l^{\prime}\right)  -\varphi\,,1\right)  $ $\left(
\ref{J-R}\right)  $, can be written in general form as
\[
H_{_{WZW}}\left(  l,\varphi\right)  =\frac{k^{2}}{2}\left(  l^{\prime}%
l^{-1},\mathbb{L}_{1}l^{\prime}l^{-1}\right)  _{L\mathfrak{d}}+\left\langle
Ad_{l^{-1}}^{LD\ast}\varphi,\mathbb{L}_{2}l^{\prime}l^{-1}\right\rangle
+\frac{1}{2}\left\langle Ad_{l^{-1}}^{LD\ast}\varphi,\mathbb{L}_{3}\bar{\psi
}\left(  Ad_{l^{-1}}^{LD\ast}\varphi\right)  \right\rangle
\]
for some linear self adjoint operators $\mathbb{L}_{i}:\mathfrak{d}%
\longrightarrow\mathfrak{d}$. The equations of motion $\left(
\ref{coll-trayect-1}\right)  $ for this Hamiltonian are
\begin{equation}%
\begin{array}
[c]{l}%
\dot{l}l^{-1}=k~\mathbb{L}_{2}l^{\prime}l^{-1}+\mathbb{L}_{3}\bar{\psi}\left(
Ad_{l^{-1}}^{LD\ast}\varphi\right) \\
\\
\dot{\varphi}=k\left(  Ad_{l^{-1}}^{LD}\left(  k\left(  \mathbb{L}%
_{1}+\mathbb{L}_{2}\right)  \left(  l^{\prime}l^{-1}\right)  +\left(
\mathbb{L}_{2}+\mathbb{L}_{3}\right)  \bar{\psi}\left(  Ad_{l^{-1}}^{LD\ast
}\varphi\right)  \right)  \right)  ^{\prime}\\
~~~~~~-k~ad_{l^{-1}l^{\prime}}^{L\mathfrak{d}\ast}\psi\left(  Ad_{l^{-1}}%
^{LD}\left(  k\mathbb{L}_{2}l^{\prime}l^{-1}+\mathbb{L}_{3}\bar{\psi}\left(
Ad_{l^{-1}}^{LD\ast}\varphi\right)  \right)  \right)
\end{array}
\label{coll-eq-motion}%
\end{equation}
When restricted to $\left(  \hat{J}^{R}\right)  ^{-1}\left(  0\,,1\right)
=\left\{  \left(  l,k\psi\left(  l^{-1}l^{\prime}\right)  \right)  \,/\,l\in
LD\right\}  \cong LD$, they become into
\[
\dot{l}l^{-1}=k\left(  \mathbb{L}_{2}+\mathbb{L}_{3}\right)  l^{\prime}l^{-1}%
\]%
\begin{align*}
\dfrac{d}{dt}\psi\left(  l^{-1}l^{\prime}\right)   &  =k\left(  Ad_{l^{-1}%
}\left(  \left(  \mathbb{L}_{1}+\mathbb{L}_{2}\right)  \left(  l^{\prime
}l^{-1}\right)  +\left(  \mathbb{L}_{2}+\mathbb{L}_{3}\right)  \left(
l^{\prime}l^{-1}\right)  \right)  \right)  ^{\prime}\\
&  -k~ad_{l^{-1}l^{\prime}}^{L\mathfrak{d}\ast}Ad_{l}^{LD\ast}\psi\left(
\left(  \mathbb{L}_{2}+\mathbb{L}_{3}\right)  l^{\prime}l^{-1}\right)
\end{align*}
Observe that for $\mathbb{L}_{1}=-\mathbb{L}_{2}$, the second equation is
derived from the first one, ensuring the Hamiltonian vector fields are tangent
to the reduced submanifold $\left(  \hat{J}^{R}\right)  ^{-1}\left(
0\,,1\right)  $. Thus, a suitable quadratic Hamiltonian for the WZW-model must
have the form%
\begin{equation}
H_{_{WZW}}\left(  l,\varphi\right)  =k\left(  \bar{\psi}\left(  Ad_{l^{-1}%
}^{LD\ast}\varphi\right)  -\dfrac{k}{2}l^{\prime}l^{-1},\mathbb{L}%
_{2}l^{\prime}l^{-1}\right)  _{L\mathfrak{d}}+\dfrac{1}{2}\left(  \bar{\psi
}\left(  Ad_{l^{-1}}^{LD\ast}\varphi\right)  ,\mathbb{L}_{3}\bar{\psi}\left(
Ad_{l^{-1}}^{LD\ast}\varphi\right)  \right)  _{L\mathfrak{d}}
\label{coll-ham-3}%
\end{equation}
that, when reduced to $\Omega D$, becomes into%
\begin{equation}
\left.  H_{_{WZW}}\left(  l,\varphi\right)  \right\vert _{\Omega D}=\dfrac
{1}{2}\left(  \bar{\psi}\left(  \hat{\Phi}\left(  \left[  l\right]  \right)
\right)  ,\left(  \mathbb{L}_{2}+\mathbb{L}_{3}\right)  \bar{\psi}\left(
\hat{\Phi}\left(  \left[  l\right]  \right)  \right)  \right)  _{L\mathfrak{d}%
} \label{coll-ham-4}%
\end{equation}
unveiling its collective form in the momentum map $\hat{\Phi}$, for a
quadratic Hamiltonian function $\mathsf{h}:\left(  L\mathfrak{d}_{\Gamma
}\right)  ^{\ast}\rightarrow\mathbb{R}$.

In order to pass to $\Omega D,$we observe that $l=l_{o}\in D$ and the reduced
equation of motion implies $\dot{l}_{o}l_{o}^{-1}=0$, so that
\[
\dfrac{d\left[  l\right]  }{dt}\left[  l\right]  ^{-1}=k~\mathbb{L~}%
\dfrac{d\left[  l\right]  }{d\sigma}\left[  l\right]  ^{-1}%
\]
with $\mathbb{L=L}_{2}+\mathbb{L}_{3}$, which is derived from $\left(
\ref{coll-trayect-1}\right)  $. Finally, it is easy to see that the
corresponding action functional is%
\begin{equation}
S_{_{WZW}}(l)=\frac{1}{2}\int_{\Sigma}\left\langle \partial_{\theta}%
ll^{-1},\partial_{t}ll^{-1}\right\rangle +\frac{1}{12}\int_{B}\left\langle
dll^{-1},[dll^{-1},dll^{-1}]\right\rangle +\frac{1}{2}\int_{\Sigma
}\left\langle \partial_{\theta}ll^{-1},\mathbb{L}\partial_{\theta}%
ll^{-1}\right\rangle \label{actionWZW}%
\end{equation}
where $\Sigma$ is a 1+1 domain with a periodic variable $\theta,$and $B$ is a
3 dimensional domain such that $\partial B=\Sigma.$ The "initial values" for
the Hamiltonian equation of motion $\left(  \ref{coll-eq-motion}\right)  $
which comes from $\left(  \ref{coll-trayect-1}\right)  $ are boundary
conditions for the fields $l(\sigma,t)$. This boundary conditions fix the
topology of $\Sigma$, the main examples are: if the condition is to be defined
for all non negative time and $l(\sigma,t=0)=e$ for all $\sigma$, then the
domain of the fields $\Sigma$ has the topology of the disc; if the condition
is to be defined for all finite time and $l(\sigma,t=0)=l_{0}(\sigma)$ for
some $l_{0}\in LD$, then the domain of the fields $\Sigma$ has the topology of
the cylinder.

Now, the first two terms on the action give the potential 1-form for the
symplectic 2-form $\omega_{\Omega D}$ in $\Omega D$ and the third term is the
corresponding hamiltonian. We recognize here the WZW model first proposed by
Klimcik and Severa if we take a specific choice of the operator $\mathbb{L}.
$Up to the moment, there are no constrains on this operators but we shall see
below that this constrains appear in order to reproduce sigma model like
lagrangians on the targets $G$ and $G^{\ast}$ and we also describe how the
boundary conditions on the fields get mapped to the dualizable subspaces.

\subsection{Hamiltonian and Lagrangian T-dual sigma models}

\bigskip As explained above, classical $T$-duality is a consequence of a
common collective dynamics on the non trivial intersection of the images of
momentum maps of systems whose phase spaces are $LD^{\wedge}$-modules. This
dynamics is fixed by a Hamiltonian function $\mathsf{h}:\left(  L\mathfrak{d}%
_{\Gamma}\right)  ^{\ast}\longrightarrow\mathbb{R}$ and the equation of motion
$\left(  \ref{coll-trayect-1}\right)  $ describes de hamiltonian vector fields
mapped by the momentum maps. Analyzing the dynamics of the WZW-model in the
previous section, we fixed the collective dynamics to be a quadratic
Hamiltonian function so that the hamiltonian function on the sigma model phase
space $T^{\ast}LG$ get fixed to be%

\[
H_{\sigma}\left(  g,\;\eta\right)  =\dfrac{1}{2}\left(  \bar{\psi}\left(
\mu\left(  g,\eta\right)  \right)  ,\mathbb{L}\bar{\psi}\left(  \mu\left(
g,\eta\right)  \right)  \right)  _{L\mathfrak{d}}%
\]
We note that only the symmetric part of the operator $\left(  \mathbb{L}%
_{2}+\mathbb{L}_{3}\right)  $ with respect to the bilinear form $\left(
,\right)  _{L\mathfrak{d}}$ contributes. So we call this symmetric part
$\mathcal{E}:\mathcal{\mathfrak{d}}\rightarrow\mathfrak{d}$ and, using eq.
$\left(  \ref{dr-LG*-E-2}\right)  $,we write
\begin{equation}
H_{\sigma}\left(  g,\;\eta\right)  =\dfrac{1}{2}\left(  Ad_{g}^{LD}%
\eta+k~g^{\prime}g^{-1},\mathcal{E}\left(  Ad_{g}^{LD}\eta+k~g^{\prime}%
g^{-1}\right)  \right)  _{L\mathfrak{d}} \label{sigma-colham-2}%
\end{equation}

In order to recover the Lagrangian functional associated to this hamiltonian,
we infer the inverse Legendre transformation from the Hamilton equation of
motion for $g$
\[
g^{-1}\dot{g}=\Pi_{\mathfrak{g}}\mathcal{E}_{g}\Pi_{\mathfrak{g}^{\ast}}%
(\eta)+\Pi_{\mathfrak{g}}\mathcal{E}_{g}\Pi_{\mathfrak{g}}(g^{-1}g^{\prime})
\]
where $\mathcal{E}_{g}=$ $Ad_{g^{-1}}^{LD}\mathcal{E}Ad_{g}^{LD}$. For
simplicity, we set from now on $k=1$ and assume the Legendre transformation is
non-singular which require the operator $\Pi_{\mathfrak{g}}\mathcal{E}_{g}%
\Pi_{\mathfrak{g}^{\ast}}:\mathfrak{g}^{\ast}\longrightarrow\mathfrak{g}$ to
be invertible for all $g\in G$. \bigskip However, it must be remarked that non
invertibility would give rise to constrains and gauge symmetries, leading to
coset sigma models \cite{Cosets} and constrained systems like the WZNW
\cite{Alek-Klim} for an appropriate choice of the kernel.

\bigskip Let us name $\mathcal{G}_{g}=(\Pi_{\mathfrak{g}}\mathcal{E}_{g}%
\Pi_{\mathfrak{g}^{\ast}})^{-1}:\mathfrak{g}\longrightarrow\mathfrak{g}^{\ast
}$ and $\mathcal{B}_{g}=-\mathcal{G}_{g}\circ\Pi_{\mathfrak{g}}\mathcal{E}%
_{g}\Pi_{\mathfrak{g}}:\mathfrak{g}\longrightarrow\mathfrak{g}^{\ast}$, then
\[
\eta=\mathcal{G}_{g}\left(  g^{-1}\dot{g}\right)  +\mathcal{B}_{g}%
(g^{-1}g^{\prime})
\]
Now a question arise: when does the resulting Lagrangian is in a sigma model
form?\ Recall that the Lagrangian of a (non singular) sigma model can be write
in the form \cite{Alvarez-npb}%
\begin{equation}
L=\left\langle g^{-1}g_{-},(\mathbf{G}_{g}+\mathbf{B}_{g})g^{-1}%
g_{+}\right\rangle \label{modsigmaG}%
\end{equation}
for $\mathbf{G}_{g}$ being a symmetric invertible operator (the metric),
$\mathbf{B}_{g}\ $an antisymmetric operator (the $B$-field), both from
$\mathfrak{g}\longrightarrow\mathfrak{g}^{\ast}$and depending on the point
$g\in G$ (the symmetry properties referred to the bilinear form given by the
pairing $\left\langle ,\right\rangle $).

The answer to this question is given by the following Lemma, in terms of the
algebraic properties of the operator $\mathcal{E}_{g}\mathcal{\ }$ in the
vector space $\mathfrak{g}\oplus\mathfrak{g}^{\ast}$

\begin{description}
\item \textbf{Lemma: }\textit{The Lagrangian coming from the collective
hamiltonian given by the symmetric operator }$\mathcal{E}$\textit{\ defines a
sigma model given by the Lagrangian}$\left(  \ref{modsigmaG}\right)
$\textit{\ with }$\mathbf{G}_{g}=\mathcal{G}_{g}$\textit{\ and }%
$\mathbf{B}_{g}=\mathcal{B}_{g}$\textit{\ iff one of the following equivalent
conditions are fulfilled for each }$G$

\begin{enumerate}
\item $\mathcal{B}_{g}$ is antisymmetric and $\mathcal{G}_{g}-\mathcal{B}%
_{g}(\mathcal{G}_{g})^{-1}\mathcal{B}_{g}=\Pi_{\mathfrak{g}^{\ast}}%
\mathcal{E}_{g}\Pi_{\mathfrak{g}}$

\item $(\mathcal{E}_{g})^{2}=1$

\item $\mathcal{E}^{2}=1$

\item As a block matrix in $\mathfrak{g}\oplus\mathfrak{g}^{\ast},$ we have%
\begin{equation}
\mathcal{E}_{g}=\left(
\begin{array}
[c]{cc}%
-(\mathcal{G}_{g})^{-1}\mathcal{B}_{g} & (\mathcal{G}_{g})^{-1}\\
\mathcal{G}_{g}-\mathcal{B}_{g}(\mathcal{G}_{g})^{-1}\mathcal{B}_{g} &
\mathcal{B}_{g}(\mathcal{G}_{g})^{-1}%
\end{array}
\right)  \label{GBmatrix}%
\end{equation}

\item $\mathfrak{d=g}\oplus\mathfrak{g}^{\ast}=\mathcal{E}_{g}^{+}%
\oplus\mathcal{E}_{g}^{-}$ where $\mathcal{E}_{g}^{\pm}$ are the $\pm1$
eigenspaces of $\mathcal{E}_{g}$ having the dimension equal to $\dim
\mathfrak{g}$ and being orthogonal to each other.
\end{enumerate}

\item \textit{Conversely, the Hamiltonian coming from a Lagrangian }$\left(
\ref{modsigmaG}\right)  $\textit{\ is in collective motion form for the moment
map }$\mu$ \textit{and quadratic non singular hamiltonian on }$L\mathfrak{d}%
$\textit{\ if the operator defined by} $\left(  \ref{GBmatrix}\right)  $
\textit{satisfies} $\mathcal{E}_{g}=$ $Ad_{g^{-1}}^{LD}\mathcal{E}_{e}%
Ad_{g}^{LD}$.\textit{\ }
\end{description}

The proof of this lemma is straightforward. It gives the exact relation
between the collective hamiltonian form and the sigma model data
\cite{KS-1}\cite{KlimYB}. We remark that the equivalences rely only on the
algebraic properties of the vector space $\mathfrak{\ g}\oplus\mathfrak{g}%
^{\ast}$ with the pairing as a bilinear form and the symmetry of $\mathcal{E}$
($\mathcal{E}_{g}$ will by also symmetric by the $Ad$-$D$ invariance of
$\left\langle ,\right\rangle ).$

Note that this kind of operators can be given by (generalized) complex
structures on the double algebra. This observation becomes relevant in the
supersymmetric case \cite{klim-park}.

In order to give a more explicit expression for the sigma model Lagrangian, we
introduce graph coordinates for the eingenspaces $\mathcal{E}_{g}^{\pm}$ on
$\mathfrak{g}\oplus\mathfrak{g}^{\ast}$(following the description of
\cite{Majid-Begg})%
\[
\mathcal{E}_{g}^{\pm}=\{X\oplus(\mathcal{B}_{g}X\pm\mathcal{G}_{g}%
X),X\in\mathfrak{g}\}
\]
This can be easily inferred from the matrix form of the operator
$\mathcal{E}_{g}.$ Now, using the dual graph coordinates%
\[
\mathcal{E}_{g}^{\pm}=\{\phi\oplus(\mathcal{B}_{g}\pm\mathcal{G}_{g})^{-1}%
\phi,\phi\in\mathfrak{g}^{\ast}\}
\]
and relating them to the ones for $g=e,$ noting that $v\in\mathcal{E}_{g}%
^{\pm}$ iff $Ad_{g}v\in\mathcal{E}_{e}^{\pm}$ so
\begin{align*}
Ad_{g^{-1}}(\phi\oplus(\mathcal{B}_{e}\pm\mathcal{G}_{e})^{-1}\phi)  &
=\Pi_{\mathfrak{g}^{\ast}}Ad_{g^{-1}}\phi\oplus Ad_{g^{-1}}((\mathcal{B}%
_{e}\pm\mathcal{G}_{e})^{-1}\phi+Ad_{g}\Pi_{\mathfrak{g}}Ad_{g^{-1}}\phi)\\
&  =\Pi_{\mathfrak{g}^{\ast}}Ad_{g^{-1}}\phi\oplus(\mathcal{B}_{g}%
\pm\mathcal{G}_{g})^{-1}(\Pi_{\mathfrak{g}^{\ast}}Ad_{g^{-1}}\phi)
\end{align*}
and we can deduce that
\[
(\mathcal{B}_{g}\pm\mathcal{G}_{g})^{-1}(\phi)=\Pi_{\mathfrak{g}}Ad_{g^{-1}%
}\Pi_{\mathfrak{g}}((\mathcal{B}_{e}\pm\mathcal{G}_{e})^{-1}\phi+Ad_{g}%
\Pi_{\mathfrak{g}}Ad_{g^{-1}}\Pi_{\mathfrak{g}^{\ast}}Ad_{g}\phi)
\]
Finally
\begin{equation}
\Pi_{\mathfrak{g}}Ad_{g}\Pi_{\mathfrak{g}}(\mathcal{B}_{g}\pm\mathcal{G}%
_{g})^{-1}\Pi_{\mathfrak{g}^{\ast}}Ad_{g}\Pi_{\mathfrak{g}^{\ast}%
}=(\mathcal{B}_{e}\pm\mathcal{G}_{e})^{-1}+\pi(g) \label{bivec-graph}%
\end{equation}
where $\pi(g)=\Pi_{\mathfrak{g}}Ad_{g}\Pi_{\mathfrak{g}^{\ast}}Ad_{g^{-1}}%
\Pi_{\mathfrak{g}^{\ast}}=-\pi^{R}(g^{-1})$, and $\pi^{R}$ gives the Poisson
bivector right translated to the origin on the Poisson-Lie group $G$ coming
from the Lie-bialgebra structure of $(\mathfrak{g},\mathfrak{g}^{\ast})$ (see
\cite{Lu-We}, for example).

So, coming back to the sigma model Lagrangian, we have%
\[
L=\left\langle g^{-1}g_{-},(\mathcal{B}_{g}+\mathcal{G}_{g})g^{-1}%
g_{+}\right\rangle =\left\langle g_{-}g^{-1},((\mathcal{B}_{e}+\mathcal{G}%
_{e})^{-1}+\pi(g))^{-1}g_{+}g^{-1}\right\rangle
\]
where in the last expression we recognize the Lagrangian of the sigma model on
the target $G$ first introduced by Klimcik and Severa \cite{KS-1}.

The corresponding dual construction can be repeated following analogous steps,
interchanging the roles of $G$ and $G^{\ast}$, yielding the dual sigma model
Lagrangian on the target $G^{\ast}$%
\[
\tilde{L}=\left\langle \tilde{g}_{-}\tilde{g}^{-1},((\mathcal{B}%
_{e}+\mathcal{G}_{e})+\tilde{\pi}(\tilde{g}))^{-1}\tilde{g}_{+}\tilde{g}%
^{-1}\right\rangle
\]
where $\tilde{\pi}$ is the corresponding Poisson bivector of the Poisson-Lie
structure on $G^{\ast}$, coming from the bialgebra $(\mathfrak{g}^{\ast
},\mathfrak{g}).$

From the construction developed on the preceding sections we know that this
two models are \emph{"dual"} to each other and to the WZW model defined by
$\left(  \ref{actionWZW}\right)  ,$ in the sense that solutions contained in
the dualizable subspace in one model can be mapped through the coadjoint orbit
$\mathcal{O}$ to the other model, and the \emph{duality transformation}
involves finding the appropriate curve in $LD^{\wedge}$ and generating the
dual flows by the action of this curve on the initial value. Conversely, we
can ask when a generic sigma model on the target $G$ will be dualizable in the
above sense. The answer to this question within the Lagrangian formalism was
given in the pioneer works \cite{KS-1}. So we conclude this section giving the
exact relation between the Lagrangian dualizability conditions (the so called
Poisson-Lie symmetry of the sigma model Lagrangian) and the information
contained in our Hamiltonian approach. To that end, following \cite{KS-1}, we
introduce the following 1-form over $\Sigma$ with values in $\mathfrak{g}%
^{\ast}$
\[
J=\Pi_{\mathfrak{g}^{\ast}}(G+B)g^{-1}g_{+}dx^{+}-\Pi_{\mathfrak{g}^{\ast}%
}(G-B)g^{-1}g_{-}dx^{-}%
\]
and we recall that a sigma model given by $\left(  \ref{modsigmaG}\right)  $
is called (right) PL-symmetric with respect to $\mathfrak{g}^{\ast}$ if
\begin{equation}
J=\frac{1}{2}[J,J]_{\mathfrak{g}^{\ast}} \label{PLsym}%
\end{equation}
over the solutions and where a Lie bracket is given on $\mathfrak{g}^{\ast}$.
It was shown that this equations require certain compatibility conditions,
namely, the bracket on $\mathfrak{g}^{\ast}$ should be such that
$(\mathfrak{g},\mathfrak{g}^{\ast})$ becomes a Lie bialgebra, and that the
$G$-dependent operators $\mathbf{G}_{g}\ $and $\mathbf{B}_{g}$ defining the
sigma model should satisfy the compatibility condition
\begin{equation}
\mathcal{L}_{X^{L}(g)}\left\langle Y,(G-B)Z\right\rangle =-\left\langle
X,ad_{\Pi_{\mathfrak{g}^{\ast}}(G-B)Z}^{\mathfrak{g}^{\ast}}(\Pi
_{\mathfrak{g}^{\ast}}(G+B)Y)\right\rangle \label{PLsymcompat}%
\end{equation}
for $X,Y,Z\in\mathfrak{g,}$ $\mathcal{L}_{X^{L}(g)}\ $is the Lie derivative
with respect to $X^{L},$ the left invariant vector field on $LG$ generated by
$X$ for all $g\in LG$. Note that if $\mathbf{G}_{g}-\mathbf{B}_{g}$ is
$G$-independent, this compatibility condition on $\mathbf{G}_{g}$ and
$\mathbf{B}_{g}$ defines a quasitriangular structure on the Lie bialgebra
$(\mathfrak{g},\mathfrak{g}^{\ast})$ \cite{Majid-Begg} \cite{KlimYB}..

This PL-symmetry condition defines, when $\Sigma$ is contractile (for example,
with the topology of the disc), a function $\tilde{h}:\Sigma\rightarrow
G^{ast}$ such that $J=d\tilde{h}\tilde{h}^{-1}$ and is easy to see that the
equation $\left(  \ref{PLsym}\right)  $ becomes equivalent to
\[
(1\pm\mathcal{E})l_{\mp}l^{-1}=0
\]
for $l=g\tilde{h}$ and $\mathcal{E}=Ad_{g}\mathcal{E}_{g}Ad_{g^{-1}}$ with
$\mathcal{E}_{g}$ the operator given by the matrix $\left(  \ref{GBmatrix}%
\right)  $ in terms of $\mathbf{G}_{g}$ and $\mathbf{B}_{g}$. Moreover, the
compatibility condition first order equation $\left(  \ref{PLsymcompat}%
\right)  $ defines how $\mathbf{G}_{g}$ and $\mathbf{B}_{g}$ depend on $g\in
G$ and it can be proved that it is equivalent to the fact that the operator
$\mathcal{E}$ just defined is constant for all $g\in G$, so $\mathcal{E}_{e}$
plays the role of initial values for these equations. We see that
$\mathcal{E}$ fulfills the conditions of the previous Lemma, so we have

\begin{description}
\item \textbf{Lemma: }\textit{Let }$\Sigma$\textit{\ be contractile and with a
periodic spatial coordinate }$\sigma$\textit{. A sigma model Lagrangian given
in the form }$\left(  \ref{modsigmaG}\right)  $\textit{\ is (right) PL
symmetric respect to }$g^{\ast}$\textit{\ iff the corresponding hamiltonian
function on }$LT^{\ast}G$\textit{\ is in collective motion form for the moment
map }$\mu$\textit{\ and the quadratic function }$\left\langle v,\mathcal{E}%
v\right\rangle $\textit{\ on }$L\mathfrak{d}_{\Gamma}^{\ast}$\textit{\ defined
by a symmetric and idempotent operator }$\mathcal{E}$\textit{\ on }%
$d$\textit{.}
\end{description}

In the case of the cylinder topology (remember the relation we stated between
the topology of the 1+1 domain and the initial values for the Hamiltonian
equations of motion for $l(\sigma,t)$) , this is also true once we have
imposed a unit monodromy constraint for the current $J$ (see below).

Finally, we shall comment on the restriction to the dualizable subspaces. Up
to now, we know by construction that there is a (unique up to constant
$G^{\ast}$ elements) correspondence between solutions of the dualizable sigma
model $(g,\tilde{h})$ and hamiltonian integral curves $(g,\tilde{h}^{\prime
}\tilde{h}^{-1})$. Now, the image of such a solution through the momentum map
$\mu$ lies in the coadjoint orbit $\mathcal{O}\left(  \tilde{h}^{\prime}%
\tilde{h}^{-1},1\right)  $ inside $\left(  L\mathfrak{d}_{\Gamma}\right)
^{\ast}$ and it is easy to see that $\mathcal{O}\left(  \tilde{h}^{\prime
}\tilde{h}^{-1},1\right)  =\mathcal{O}$ iff $\tilde{h}$ is periodic in the
spatial variable (i.e., iff it is a loop for all t). So the restriction to the
dualizable subspace can be expressed as a unit monodromy constrain on
$J=d\tilde{h}\tilde{h}^{-1}$,
\[
\tilde{h}(0,t)=\tilde{h}(2\pi,t)
\]
or equivalently
\[
P\int_{\gamma}d\tilde{h}\tilde{h}^{-1}=\tilde{e}%
\]
for any closed path $\gamma$ homotopic to a constant time loop in $\Sigma$,
and the same holds for the dual model, as first noted by Klimcik and Severa.

So, as the Lagrangian on the $G$ target describes the Hamiltonian dynamics in
the whole phase space $LT^{\ast}G$ it is natural to ask what kind of models
arises when we replace $LT^{\ast}G^{\ast}$ by other phase-space such that it
has a non-trivial intersection with the other coadjoint orbits (the image of
the subspaces restricted by non-trivial monodromy conditions) $\mathcal{O}%
\left(  \alpha,1\right)  ,$ with $\alpha\in L\mathfrak{g}^{\ast}, $ which also
lie in $\mu(LT^{\ast}G).$ Such models should have phase spaces consisting of
non-closed paths in $G^{\ast}$(because of the non-trivial monodromy of
$\alpha)$. Examples of this models are the ones given in \cite{Monodromic}.
\bigskip

The reader might also note that the Poisson structures on such open path
spaces are closely related to the ones associated to the chirally extended
WZW\ phase space \cite{Feheretal}, which are also $LD$ spaces, and from this
point of view one could have a better understanding of the appearance of
(finite dimensional) P-L symmetries generated by the monodromy matrix of the
resulting open strings variables \cite{KSmoment}.

\bigskip

\section{V- Examples}

We will now give some examples to illustrate on the construction of the
duality transformations and admissible subspaces for special simple choices of
the double group and the Hamiltonian dynamics, recovering a full explicit
description of known results on target space duality.

\subsection{Abelian duality and $R\longrightarrow\frac{1}{R}$ momentum-winding
interchange.}

In this example, we take a trivial Lie bialgebra $(\mathfrak{g},[,]=0,$
$\delta=0),$ which has a trivial dual bialgebra and a trivial double
$(\mathfrak{g}^{\ast},[,]=0,\delta=0),$ $(\mathfrak{d},[,]=0,\delta=0)$
respectively. Moreover, we take $\mathfrak{g}$ to be the Lie algebra of the
1-dimensional abelian group $G=U_{R}(1)$ thought as a circle of radius $R$
with group law the translation along the circle. Being the bilinear form
$\left(  ,\right)  _{\mathfrak{d}}$ on the double Lie algebra the pairing
between $\mathfrak{g}$ and its dual $\mathfrak{g}^{\ast},$ we can choose the
dual group to be $G^{\ast}=U_{\frac{1}{R}}(1),$ the \emph{dual} circle of
radius $\frac{1}{R},$ since we can naturally think of $\mathfrak{g}$ and its
dual $\mathfrak{g}^{\ast}$ as the corresponding tangent spaces at the origin
and if we parametrize the group elements as $Rx$ and $R^{-1}x$ with
$x\in\lbrack0,2\pi]$ then $\left\langle R\partial_{x},R^{-1}d_{x}\right\rangle
=1.$ We choose the (non-simply connected) double group to be $D=U_{R}(1)\times
U_{R^{-1}}(1)$. The Poisson bracket on $L\mathfrak{d}^{\wedge\ast}$ is pure
central extension $\{,\}\equiv\Gamma,$ and we choose the Hamiltonian function
on $L\mathfrak{d}^{\wedge\ast}$ to be
\[
\mathfrak{H}(X,\xi,a)\mathfrak{=}\int_{S^{1}}d\sigma\;\left(  \dfrac{1}%
{2R^{2}}\xi^{2}+\dfrac{R^{2}}{2}X^{2}\right)
\]

The phase-spaces of the dual models are $LT^{\ast}U_{R}(1)$ and $LT^{\ast
}U_{\frac{1}{R}}(1),$ with parametrized elements $\gamma=(\theta(\sigma
),\pi(\sigma))\in LT^{\ast}U_{R}(1)$ and $\tilde{\gamma}=(\tilde{\theta
}(\sigma),\tilde{\pi}(\sigma))\in LT^{\ast}U_{\frac{1}{R}}(1)$. The moment
maps associated with the $LD$ action are $\mu(\theta(\sigma),\pi
(\sigma))=(\dfrac{d}{d\sigma}\theta(\sigma)+\pi(\sigma),1)$ and $\tilde{\mu
}(\tilde{\theta}(\sigma),\widetilde{\pi}(\sigma))=(\dfrac{d}{d\sigma}%
\tilde{\theta}(\sigma)+\tilde{\pi}(\sigma),1).$ The Hamiltonians on the
phase-spaces written on the collective form are
\[
H(\gamma;\sigma)=\int_{S^{1}}d\sigma\;[\dfrac{1}{2R^{2}}\pi(\sigma)^{2}%
+\dfrac{R^{2}}{2}(\dfrac{d\theta}{d\sigma})^{2}]
\]%
\[
\tilde{H}(\tilde{\gamma};\sigma)=\int_{S^{1}}d\sigma\;[\dfrac{1}{2R^{2}%
}(\dfrac{d\tilde{\theta}}{d\sigma})^{2}+\dfrac{R^{2}}{2}\tilde{\pi}%
(\sigma)^{2}]
\]

The duality transformation $\Psi:\mu^{-1}\left(  \mathcal{O}\right)
\longrightarrow\tilde{\mu}^{-1}\left(  \mathcal{O}\right)  $ can be
constructed following the arrows of the diagram $(\ref{T-duality-1})$%
\begin{equation}
\begin{diagram}[h=1.9em] LT^{\ast }U_{R}(1)&&&&LT^{\ast }U_{1/R}(1)\\ &&&&\\ (\theta (\sigma ),\pi (\sigma )) &\rTo^{\mu}& ( \theta ^{\prime }(\sigma ),\pi (\sigma ),1) &\lDashto^{\tilde{\mu}} &({\tilde \theta}_o+\int_{0}^{\sigma }\pi (\zeta )d\zeta \; ,\;\dfrac{d\theta }{d\sigma }(\sigma ))\\ &&&&\\ &&\dTo^{{\hat{\Phi}}^{-1}} &\ruTo_{{\tilde\varsigma}_{{\tilde \theta}_o}}&\\ &&&&\\ &&\theta (\sigma ) - \theta (0) \times \int_{0}^{\sigma }d\zeta \;\pi (\zeta )&&\\ \end{diagram}
\end{equation}
so we get
\[
\Psi(\gamma(\sigma))=\tilde{\gamma}(\sigma)=(\int_{0}^{\sigma}\pi(\zeta
)d\zeta,\;\dfrac{d\theta}{d\sigma}(\sigma))
\]
Proceeding analogously, starting from the dual part, we obtain $\tilde{\Psi}.
$ By construction
\[
\Psi^{\ast}\tilde{H}=H
\]

The admissible subspace $\mu^{-1}\left(  \mathcal{O}\right)  $ in $LT^{\ast
}U_{R}(1)$ is of the form $\{(\theta,\tilde{\alpha})\bullet(\theta
_{0},0)=(\theta+\theta_{0},\tilde{\alpha}^{\;\prime})$ $/:(\theta
,\tilde{\alpha})\in\Omega D\}$ since $D$ is abelian and so the dressing
actions are trivial. Similarly, $\tilde{\mu}^{-1}\left(  \mathcal{O}\right)
=\{(\tilde{\theta},\alpha)\bullet(\tilde{\theta}_{0},0)=(\tilde{\theta}%
+\tilde{\theta}_{0},\alpha^{\;\prime})/:(\tilde{\theta},\alpha)\in\Omega D\}$
in $LT^{\ast}U_{\frac{1}{R}}(1)$. We note that the elements in $\Omega D$
giving the duality transformations have $\theta(0)=0$ or they have no
\emph{momentum zero modes} in their Fourier expansion. This corresponds, as in
the general PL case, to the unit monodromy constraint (see \cite{Monodromic}).
Now, the topology of the $U(1)$ targets allows us to introduce a refined
description of the dualizable subspaces.

As every element $(\theta,\tilde{\theta})\in L(U_{R}(1)\times U_{\frac{1}{R}%
}(1))$ is classified by its homotopic class or \textquotedblright winding
number\textquotedblright\ we then define the subsets
\[
L(n,m)=\{(\theta,\tilde{\theta})\in L(U_{R}(1)\times U_{\frac{1}{R}%
}(1))/:\;\deg\theta=n\;and\;\deg\tilde{\theta}=m\}\;
\]

\medskip So $\mu^{-1}\left(  \mathcal{O}\right)  =\bigcup(L_{R}(n,m))=\bigcup
\{(\theta,\pi)/:\deg\theta=n\;and\;\int_{S^{1}}\pi=2\pi m\}$ and $\tilde{\mu
}^{-1}\left(  \mathcal{O}\right)  =\bigcup(L_{\frac{1}{R}}(n,m))=\bigcup
\{(\tilde{\theta},\widetilde{\pi})/:\deg\tilde{\theta}=n\;and\;\int_{S^{1}%
}\widetilde{\pi}=2\pi m\}.$ Moreover, its easy to see that the Hamiltonian
flows preserves these winding numbers and so the $L_{R}(n,m)$ and $L_{\frac
{1}{R}}(n,m)$ are sub-Hamiltonian systems of $LT^{\ast}U_{R}(1)$ and
$LT^{\ast}U_{\frac{1}{R}}(1)$ respectively, which lie inside the dualizable
subspaces. Finally, we see that the duality transformation $\Psi$ maps
$L_{R}(n,m)$ to $L_{\frac{1}{R}}(m,n)$ interchanging $R\longrightarrow R^{-1}$
and the winding number $n$ to be the momentum number in the dual model and the
momentum number $m$ to the winding number in the dual model. Hence we have
recovered the momentum-winding duality transformation and the domain of these
transformation in the phase spaces as described in \cite{Alvarez-Liu} within
our general framework.

\bigskip

\subsection{Semi-abelian or non-abelian $G\longleftrightarrow g^{\ast} $
duality}

In this example, we take the bialgebra $\mathfrak{g}$ to be semi-trivial, that
is, $(\mathfrak{g,}$ $[,],$ $\delta=0).$ So $(\mathfrak{g}^{\ast}%
,[,]=0,\delta)$ and the double $(\mathfrak{d},$ $[,],\delta)$ can be
identified as a Lie algebra with the semi-direct product of $\left(
\mathfrak{g,}[,]\right)  $ and $(\mathfrak{g}^{\ast},[,]=0)$ where
$\mathfrak{g}$ acts on $\mathfrak{g}^{\ast}$ by the coadjoint representation.
We will take $G$ as a compact simple group and in role of $G^{\ast}$ we
consider $\mathfrak{g}^{\ast}$ as an additive Abelian group, so the double
group can be identified with the semidirect product $D=G\rhd$ $\mathfrak{g}%
^{\ast},$ with $G$ acting by the coadjoint representation. The phase-spaces
are $LT^{\ast}G$ and $LT^{\ast}\mathfrak{g}^{\ast}$ and taking the product on
$G\rhd\mathfrak{g}^{\ast}$ to be
\[
(g,\eta)\cdot(h,\lambda)=\left(  gh,\lambda+Ad_{h}^{LG\ast}\eta\right)
\]
we have that $(g,\eta)^{-1}=(g^{-1},-Ad_{g^{-1}}^{LG\ast}\eta)$ and the
momentum maps corresponding to the $LD$ actions on them are $\mu
(g,\eta)=(k~g^{\prime}g^{-1}+Ad_{g^{-1}}^{LG\ast}\eta,1)$ for $(g,\eta)\in
L(G\times\mathfrak{g}^{\ast})$, the left trivialization of $LT^{\ast}G$,
and$\;\tilde{\mu}(\eta,X)=(X+\eta^{\prime}+ad_{X}^{LG\ast}\eta,1\;)$ for
$(\eta,X)\in$ $LT^{\ast}\mathfrak{g}^{\ast}$. The coadjoint $D\ $and
$LD\ $actions become%
\[
Ad_{(g,\eta)}^{D}(X,\xi)=(Ad_{g}^{G}X,\ Ad_{g^{-1}}^{G\ast}\xi+Ad_{g^{-1}%
}^{G\ast}ad_{X}^{G\ast}\eta)
\]%
\[
\widehat{Ad}_{(g,\eta)}^{LD\ast}(0,1)=\left(  (g,\eta)^{\prime}\cdot
(g,\eta)^{-1},1\right)  =(g^{\prime}g^{-1}+Ad_{g^{-1}}^{LG\ast}\eta^{\prime
},1)
\]
for $(g,\eta)\in LD$ acting on $L\mathfrak{d}^{\wedge\ast}$. The actions on
the cotangent bundles can be derived from the dressing action for this
particular case%
\begin{align*}
(g,\eta)  &  =(g,0)\cdot(e,\eta)=(e,\alpha)\cdot(g,0)\quad\Longrightarrow\quad
g^{\eta}=g\quad,\quad\alpha^{g}=Ad_{g}^{LG\ast}\alpha\\
(g,\eta)  &  =(e,Ad_{g^{-1}}^{LG\ast}\eta)\cdot(g,0)=(g,0)\cdot(e,\eta
)\quad\Longrightarrow\quad\eta_{g}=Ad_{g^{-1}}^{LG\ast}\eta\quad,\quad
g_{\eta}=g
\end{align*}
so that%
\begin{align*}
\mathsf{\hat{d}}\left(  (h,\xi),\left(  g,\eta\right)  \right)   &
=\mathsf{\hat{d}}\left(  (h,0)\cdot(e,\xi),\left(  g,\eta\right)  \right)
=\left(  hg,\eta+\left(  Ad_{g}^{LG\ast}\xi\right)  ^{\prime}\right) \\
\mathsf{\hat{b}}\left(  (h,\xi),\left(  \eta,X\right)  \right)   &
\equiv\mathsf{\hat{b}}\left(  (e,Ad_{h^{-1}}^{LG\ast}\xi)\cdot(h,0),\left(
\eta,X\right)  \right)  =\left(  Ad_{h^{-1}}^{LG\ast}\xi+Ad_{h^{-1}}^{LG\ast
}\eta,Ad_{h}^{G}X+h^{\prime}h^{-1}\right)
\end{align*}

From here we can get the elements of $\mu^{-1}(\mathcal{O})$ and $\tilde{\mu
}^{-1}(\mathcal{O})$%
\[
\mu^{-1}(\mathcal{O})=\mathsf{\hat{d}}\left(  (h,\xi),\left(  g_{o},0\right)
\right)  =\left(  hg_{o},Ad_{g_{o}}^{LG\ast}\xi^{\prime}\right)
\]%
\[
\tilde{\mu}^{-1}(\mathcal{O})=\mathsf{\hat{b}}\left(  (h,\xi),\left(  \eta
_{0},0\right)  \right)  =\left(  Ad_{h^{-1}}^{LG\ast}\xi+Ad_{h^{-1}}^{LG\ast
}\eta_{0},h^{\prime}h^{-1}\right)
\]
for $(h,\xi)(\sigma=0)=(e,0)$ and, following the arrows in the diagram
$\left(  \ref{T-duality-4}\right)  $,%

\[
\begin{diagram}[h=1.9em]
\left( hg_{o},Ad_{g_{o}^{-1}}^{LG\ast } \xi ^{\prime } \right) &&\rTo^{\mu}&& \widehat{Ad}_{(h,\xi )}^{LD\ast }(0,1) &&\lDashto^{{\tilde{\mu}}}&& \left( Ad_{h^{-1}}^{LG\ast } (\xi +\eta_o) ,h^{\prime }h^{-1}\right)\\
&&&&&&&&\\ &&&&\dTo^{{\hat{\Phi}}^{-1}}&&\ruTo_{{\tilde\varsigma}_{\eta_o}}&&\\
&&&&&&&&\\ &&&&\left[ (h,\xi )\right]&&&&\\ \end{diagram}
\]
we get the duality transformation%

\begin{equation}
\Psi_{\eta_{o}}\left(  hg_{o},Ad_{g_{o}^{-1}}^{LG\ast}\xi^{\prime}\right)
=\left(  Ad_{h^{-1}}^{LG\ast}\left(  \xi-\xi_{o}\right)  +Ad_{(hh_{o}%
^{-1})^{-1}}^{LG\ast}\eta_{o},h^{\prime}h^{-1}\right)  \label{td-Gxg-1}%
\end{equation}

The admissible subspace $\mu^{-1}\left(  \mathcal{O}\right)  $ in $LT^{\ast}G$
is
\[
LD\cdot(e,0)\sim(LG,\Omega\mathfrak{g}^{\ast})
\]
and $\tilde{\mu}^{-1}\left(  \mathcal{O}\right)  $ in $LT^{\ast}%
\mathfrak{g}^{\ast}$ is
\[
LD\cdot(0,0)\sim(L\mathfrak{g}^{\ast},\Omega G)
\]

The inverse duality transformation can be computed following the arrows in
$\left(  \ref{T-duality-1}\right)  $ the other direction%

\[
\begin{diagram}[h=1.9em] \left( hh_{o}^{-1}g_{o},Ad_{g_{o}}^{LG\ast }\xi ^{\prime } \right) && \rDashto^{\mu}&& \widehat{Ad}_{(h,\xi)}^{LD\ast }(0,1) &\lTo^{{\tilde{\mu}}}&\left(Ad_{h^{-1}}^{LG\ast }(\xi+\eta_o) ,h^{\prime }h^{-1}\right)\\ &&&&&&\\ &&\luTo_{\varsigma _{g_{o}}}&&\dTo_{\hat{\Phi}^{-1}}&&\\ &&&&&&\\ &&&&\left[ (h,\xi )\right] &&\\ \end{diagram}
\]
where $\left(  \eta_{o},0\right)  \in\tilde{\mu}^{-1}\left(  \mathcal{O}%
\right)  $, so we get%
\begin{equation}
\tilde{\Psi}_{g_{o}}\left(  Ad_{h^{-1}}^{LG\ast}\xi+Ad_{h^{-1}}^{LG\ast}%
\eta_{o},h^{\prime}h^{-1}\right)  =\left(  hh_{o}^{-1}g_{o},Ad_{g_{o}}%
^{LG\ast}\xi^{\prime}\right)  \label{td-Gxg-2}%
\end{equation}

This duality transformations can be obtained from a generating functional
\[
\Gamma(h,\xi)=-\int_{S^{1}}\left\langle \xi(\sigma),\;h^{\prime}%
h^{-1})\right\rangle =-\int_{S^{1}}l^{\ast}\vartheta
\]
as it is done in \cite{Alvarez-Liu}. The last equality follows for $l=(h,\xi)$
from the general formula for the generating functional of the duality
transformations and the fact that for $D=G\rhd$ $\mathfrak{g}^{\ast}$ the
symplectic form on the Double is exact $\omega^{STS}=d\vartheta$. In ref.
\cite{Alvarez-Liu}, they take a slightly different functional $\Gamma
(\varphi^{-1}\varphi^{\prime},\xi)=-\int_{S^{1}}\left\langle \xi
(\sigma),\;\varphi^{-1}\varphi^{\prime})\right\rangle $ that leads to
equivalent duality transformations which can be derived within our framework
by taking $LT^{\ast}G\sim L(G\times\mathfrak{g}^{\ast})$ trivialized by right
translations and the cocycle $\Gamma$ on $LD$ by $\Gamma(l)=l^{-1}l^{\prime}$
and repeat the whole procedure of constructing the maps $\mu$ and $\phi$ in an
analogous way changing left by right invariants. The discussion about the
right domain for the duality transformations is, within our framework, very
simple because by construction we know that the correct domains are given by
the dualizable subspaces which we explicitly constructed and, in addition, we
know that they will be invariant under any collective hamiltonian flow on the phase-spaces.

\bigskip

\section*{Conclusions}

We have analyzed some relevant geometric properties of the loop spaces related
to Poisson-Lie T-Duality, mainly centred on loop actions of the $\sigma
$-models derived from the dressing transformation lifted to the cotangent
bundle, with associated equivariant momentum maps. This allowed us to describe
and understand many of the various aspects of this duality under the
Hamiltonian formalism, like the explicit procedure of duality transformations,
a precise identification of the dualizable subspaces and their relevance, and
to reconstruct in a systematic way the well-known T-dual sigma model
Lagrangians for suitable choices of the corresponding collective Hamiltonian
dynamics. Moreover, this description allows to identify the relevant
properties of the underlying information given by the models in order to be
T-dual. In that way, we observed the same construction can reproduce known
generalizations of PLTD as coset model dualities \cite{Cosets}, duality for
matched pairs \cite{Majid-Begg}, PLT-plurality \cite{Plurality} for different
decompositions of the Drinfeld double, Buscher%
\'{}%
s duality \cite{Majid-Begg}\cite{KS-1} and duality for monodromic strings
models \cite{Monodromic}, because the underlying loop geometry enjoys the same
properties for duality as the standard case. We believe that this approach can
be generalized and adapted to cover many different (new) types of dualities
becoming a useful geometrical approach for the study of T-dualities. For
example, one can replace $LT^{\ast}G^{\ast}$ by other Hamiltonian $LD_{\Gamma
}$-space like the ones related to dual symmetric spaces \cite{SymmSpac} and,
repeating the above construction, generate a new collection of T-dual models
for each collective Hamiltonian choice. In addition, the construction itself
will give the properties of the resulting models with respect to duality. This
can also be applied for non-perfect doubles using the symplectic leaves
decomposition of \cite{Alek-Malkin}. More generally, one can build up new
types duality diagrams by considering symplectic groupoid or Poisson-Lie group
actions instead of the usual hamiltonian ones. For finite dimensional PL
actions, the construction given in this paper can be adapted to describe the
duality observed between $G$ and $G^{\ast}$ Poisson-sigma models
\cite{Poissonsigma} with its corresponding boundary-bulk duality
transformation. It would be also interesting to repeat the construction for
the chirally extended WZNW spaces \cite{Feheretal} which is both a Loop space
and a finite dimensional PL space, and relate this actions to more general
ones by the (Morita) equivalences of \cite{Xu}. In the infinite dimensional PL
case, we think that the relevant loop spaces for repeating the diagram
construction should be closely related to the ones investigated in
\cite{klim-qt}. A more general approach based on groupoid actions might be
used to study global properties of the dualities given in \cite{Alvarez-npb}.

Finally, we hope that, as this is an explicit description of T-dualities in
the Hamiltonian formalism, it will turn out to be usefull to analize the
resulting quantum T-dualities under a quantization scheme adapted to the
underlying geometry of the dual phase-spaces.

\subsection*{Acknowledgments}

H.M. thanks to CONICET for financial support. A.C. thanks to CNEA for
financial support during the Master in Physics programe at the Balseiro Institute.


\end{document}